\documentstyle[preprint,aps,eqsecnum,psfig]{revtex}
 
\def\be{\begin{equation}}
\def\ee{\end{equation}}
\def\bea{\begin{eqnarray}} 
\def\eea{\end{eqnarray}}
\def\line{\hbox to \hsize}    
\def\frac #1#2{{#1\over #2}}
\def\Tr{{\rm  Tr\,}}
\def\tr{{\rm  tr\,}}

\def\bzeta{{\overline\zeta}}
\def\bareta{{\overline\eta}}
\def\bxi{{\overline\xi}}
\def \z{{\overline z}}

\def\hH{{\hat H}}

\def\det{{\rm det\,}}
\def\Det{{\rm Det\,}}
 
\def \B{{\overline B}}

\def \ket #1{{\vert #1\rangle}}
\def \bra #1{{\langle #1\vert}}
\def \brak #1#2{{\langle#1\vert#2\rangle}}
\def\eval #1#2#3{{\langle#1\vert#2\vert#3\rangle}} 
\def\vev #1{{\langle #1\rangle}}
\def\1{\mbox{\bf 1}} 
\def\jtil{{\tilde j}}
\def\om{\omega}
\def\ptl{\partial}
\def\tom{\tilde\om}

\begin{document}
\draft %(only for revtex) 

\title{
The Semiclassical Propagator for Spin Coherent States 
}

\author{MICHAEL STONE}
\address{University of Illinois, Department of Physics\\ 1110 W. Green St.\\
Urbana, IL 61801 USA
%\\E-mail: m-stone5@uiuc.edu
}

\author{KEE-SU PARK}
\address{University of Illinois, Department of Physics\\ 1110 W. Green St.\\
Urbana, IL 61801 USA
%\\E-mail: k-park4@uiuc.edu
}

\author{ANUPAM GARG}
\address{ Northwestern
University, Department of Physics and Astronomy\\Evanston,
IL 60208 USA
%\\E-mail agarg@nwu.edu
}

\maketitle

\begin{abstract}

We  use a  continuous-time  path integral to obtain   the 
semiclassical propagator for minimal-spread  spin coherent
states. We pay particular attention to   the ``extra
phase'' discovered by Solari and Kochetov, and show that
this  correction is  related to an anomaly in the
fluctuation determinant.  We show that, once this extra
factor  is included, the   semiclassical propagator  has
the  correct short time  behaviour to  $O(T^2)$,  and 
demonstrate its consistency  under dissection of the path. 

\end{abstract}

%\newpage 

\section{Introduction}

Coherent-state path integrals for  spin were introduced by
Klauder\cite{klauder79}, and by Kuratsuji and Suzuki
\cite{kuratsuji80}. Related phase space path integrals were
introduced by Jevicki and Papanicolaou\cite{jevicki79}, and
by Nielsen  and R\"ohrlich\cite{rohrlich88}.  For a review
see \cite{klauder78,inomata92}. These path integrals have
attracted attention in  connection with geometric
quantization\cite{woodhouse}, and for providing examples
hinting at possible infinite-dimensional extensions of the
Duistermat-Heckman theorem\cite{duistermat82} on conditions
for the exactness of the stationary phase
approximation\cite{stone89,semenoff91}. Perhaps their most
significant  practical applications, however,  have been in
computations of spin tunnelling in the semiclassical
limit.  Here the  spin path-integral formalism gives a
good  qualitative description of the tunneling
process\cite{chudnovsky88,loss92,henley92}, including the
simplest and most vivid picture of the topological
quenching of spin tunneling\cite{garg93} that has recently
been seen in the magnetic molecule
Fe$_8$\cite{wernsdorfer99}. When we require precise
quantitative results, however,  the spin coherent-state
path integral runs into problems:  A straight forward
application of instanton methods to compute the tunnel
splitting\cite{kuratsuji81,garg_kim92}  yields answers that
are incorrect beyond the leading exponential
order\cite{enz86}.  A  full derivation of the splitting,
including the correct prefactor, has only recently  been
provided  by Belinicher, Providencia and
Providencia\cite{belinicher97}. These authors  showed that
the continuum limit of the discrete path integral is rather
delicate, and  in their computation the simplicity  of the
instanton method is lost.  These difficulties have  lead to
the spin path integral acquiring   a reputation  for being
unreliable---or, even worse, being meaningful only in its
discrete-time form\cite{shibata99}.   Many workers in the
field have sought alternatives to path integrals such as
discrete WKB methods \cite{hemmen86,garg99,villain00}.

This paper is  intended to effect a  rehabilitation of  the
continuous-time spin coherent-state path integral.  We
advertise and explain the origin of a previously discovered,
but largely unknown, correction to the na{\"\i}ve  form of
the semiclassical propagator. This ``extra phase''  was 
obtained by Solari\cite{solari87} as a result of a careful 
evaluation of  the discrete path form of the
path integral. It also appears, as a product of a
manipulation, apparently carried out  for convenience, in a
paper  by  Kochetov\cite{kochetov95}. We derive it here by
pointing  out that   the  functional determinant resulting
from the fluctuation integral about the classical path
possesses an anomaly.  Regulating the determinant in a
manner consistent with the underlying causal structure
leads to the extra contribution.  

The structure of the paper is as follows: In section two
we  review spin coherent states built on highest- or
lowest-weight spin-$j$ states. We focus primarily on  their
holomorphic properties. In section three we review the
properties of the classical action that appears in the path
integral for spin, stressing the importance of  boundary
terms in avoiding the over-determination problem. In
section four we compute the gaussian integral over  small
fluctuations about the classical path, and obtain the
extra-phase correction. In section five we verify that,
once  the extra contribution is  taken into account,   the
semiclassical propagator has the correct short time
behaviour. This verification is immediate  at first order
in $T$, but the agreement between our expression and the
exact result at $O(T^2)$ provides a significant test of the
correctness of our result.    
In section six  we check the
consistency of the expression for the  propagator under the
dissection of the path. We find that our semiclassical
propagator  does {\it not\/} pass this test unless we
repartition terms  between the exponent and the prefactor.
This forces us to  regard the large parameter in the
semiclassical expansion as being $j+1/2$, rather than $j$.
As a byproduct, this observation  resolves the mystery of
the divergent normalization factor that appears in most
treatments of the semiclassical propagator.  Finally, in
section seven, we compute the semiclassical propagator for
the hamiltonian $\hH= \nu J^2_z$. We confirm that our
expression obtains  the correct leading and next to leading
terms in the large-$j$ expansion.

\section{Spin Coherent States}

We  define a  family of spin coherent
states\cite{perelomov86} by
\be
\ket{z} = \exp(z \hat J_+) \ket{j,-j}.
\ee
These states are  not normalized, but  have the advantage
of being holomorphic in the parameter $z$. Consequently, matrix
elements such as $\eval{z'}{\hat O}{z}$ will be holomorphic functions
of the variable $z$, and anti-holomorphic
functions of the variable $z'$.

The inner product of two of these  states is  
\be
\brak{ z'}{z} = (1+ \z' z)^{2j},
\ee
and the left eigenstates $\bra{j,m}$  of $\hat J^2$ and $\hat J_3$ have
coherent-state
wavefunctions
\be
\psi_m^{(1)}(z)\equiv \brak{j,m}{z}= \sqrt{\frac{2j!}{(j-m)!(j+m)!}}\, z^{j+m}.
\ee
This means that a general element of the spin-$j$ Hilbert space may be
represented by a  
polynomial in $z$ of degree $n \le 2j$.

As with any family of generalized coherent states derived
from a unitary   irreducible representation of a  compact
group, Shur's lemma provides us with an overcompleteness
relation. In the present case this reads  
\be
\1 = \frac {2j+1}{\pi} \int \frac{d^2z}{(1+\z z)^{2j+2}}
\,{\ket{z}} {\bra{z}}.\
\label{EQ:overcompleteness}
\ee
Here  $2j+1$ appears because it is the dimension of the
representation. The symbol $d^2z$ is shorthand for  $dx\,dy$, and the
factor  $1/(1+\z z)^2$ combines with this to make the
invariant measure on the coset $SU(2)/U(1)$. This coset is,
of course, the two-sphere, $S^2$, equipped with 
stereographic coordinates. The south pole, corresponding to
spin down,   is at $z=0$, while the  north pole, spin up,
is at $z=\infty$ --- the one-point  compactification of the
complex plane.     The remaining factor in the measure,
$1/(1+\z z)^{2j}$, serves to normalize the  states.

%The overcompleteness provides the inner product for the
%coherent-state 
%wavefunctions.
%If $f(z) =\brak{f}{z}$ and $g(z)=\brak{g}{z}$ then
%\bea
%\brak{f}{g} &=&  
%\frac {2j+1}{\pi} \int \frac{d^2z}{(1+\z z)^{2j+2}}\,\brak{f}{z}\brak{z}{g}
%\nonumber\\&=& 
%\frac {2j+1}{\pi} \int \frac{d^2z}{(1+\z z)^{2j+2}}\,\overline{
%g(z)}f(z).
%\eea

The wavefunctions  $\psi_m^{(1)}(z)$ are  singular at the
north pole, $z=\infty$.  Indeed there is no actual 
state $\ket{\infty}$ because the phase of this putative  limiting
state would depend on the direction from which we approach
the  point at
infinity. We may, however, define a second
family of states
\be
\ket{z}_2 = \exp(z \hat J_-) \ket{j,j},
\ee
and form the wavefunctions 
\be
\psi_m^{(2)}(z)=\brak{j,m}{z}_2.
\ee
These states and wavefunctions are well defined in the vicinity of 
the north pole, but singular
near the south pole.

To find the relation between $\psi^{(2)}(z)$ and $\psi^{(1)}(z)$ we
note that the matrix identity
\be
\left[\matrix{ 1 & z \cr 0 & 1}\right]\left[\matrix{ 0 & 1 \cr -1 & 0}\right] 
=\left[\matrix{ 1 & 0 \cr z^{-1}  & 1}\right]
\left[\matrix{ -z & 0 \cr 0 & -z^{-1}}\right]
\left[\matrix{ 1 & -z^{-1} \cr 0 & 1}\right],
\ee
coupled  with the faithfulness of the spin-$\frac 12$
representation of $SU(2)$, 
implies the relation
\be
\exp({z\hat J_+})\hat w= 
\exp{(z^{-1} \hat J_-)}(-z)^{2\hat J_3} \exp{(-z^{-1} \hat J_+)},
\ee
where
$
\hat w= \exp(i\pi\hat J_2)
$
is the generator  of the Weyl group of $SU(2)$. We also note that
\be
\hat w\ket{j,j} = (-1)^{2j}\ket{j,-j},\quad \hat w\ket{j,-j} = \ket{j,j}.
\ee
Thus,
\bea
\psi_m^{(1)}(z)&=& \eval{j,m} {e^{z\hat J_+}}{j,-j}\nonumber\\
             &=& (-1)^{2j}\eval{j,m} {e^{z\hat J_+}\hat w}{j,j}\nonumber\\
             &=& (-1)^{2j}\eval{j,m} {e^{z^{-1} \hat J_-}(-z)^{2\hat J_3}
                 e^{-z^{-1} \hat J_+}}{j,j}\nonumber\\
             &=& (-1)^{2j}(-z)^{2j}\eval{j,m}
                {e^{z^{-1}\hat J_-}}{j,j}\nonumber\\
             &=& z^{2j} \psi_m^{(2)}(z^{-1}).
\eea
The coherent-state wavefunctions  $\psi_m^{(1)}$ and
$\psi_m^{(2)}$ may therefore be regarded as composing a
single  global section, $\psi_m$, of a holomorphic line
bundle with transition function $z^{2j}$ relating its
components $\psi_m^{(1)}(z)$, and
$\psi_m^{(2)}(\zeta\equiv 1/z)$ in the two coordinate
patches. It is the requirement that the transition function
and its inverse be holomorphic and single valued in the
overlap of the coordinate patches  that forces $2j$ to be
an integer.  In the sequel, all coherent states, unless
otherwise specified, will drawn from the first family,
$\ket{z}$.

The above   construction is an example of the Borel-Weil
realization of representations of compact groups as
sections of holomorphic bundles\cite{borel}.  It serves as
the paradigm for the more general theory of geometric
quantization\cite{woodhouse,kirillov}.  Because global
analyticity is characteristic of  the minimal-spread 
coherent states built on highest- (or lowest-) weight
states, and also serves (via the transition function) to
specify the Hilbert space, it is a property that should be
maintained order-by-order in any approximation scheme. 

For physical interpretations we must normalize the
coherent states. This we do by multiplying  them by 
\be
N(\z,z)= (1+\z z)^{-j}.
\ee
For example,
\be
N^2 \eval{z}{\hat J_3}{z}= j \frac {\z z-1}{\z z+1}, \quad\hbox{\rm and}\quad 
N^2 \eval{z}{\hat J_+}{z}=  \frac {2j\z}{\z z+1}.
\ee
If we recall the connection between stereographic
and spherical polar coordinates,
\be
z= e^{-i\phi}\cot \frac{\theta}{2},  
\ee
we see that
\be
j \frac {\z z-1}{\z z+1}= j\cos\theta, \quad\hbox{\rm and}\quad 
\frac {2jz}{\z z+1} =je^{-i\phi}\sin\theta.
\ee

We also note that
\bea
N^2 \eval{z}{\hat J_3^2}{z}&=& j(j-\frac 12) 
\left(\frac {\z z-1}{\z z+1}\right)^2
+\frac 12 j\nonumber\\
& =& j(j-\frac 12)\cos^2\theta + \frac j2.
\eea 
Similarly\cite{lieb73}
\bea
N^2 \eval{z}{\hat J_1^2}{z}&=& j(j-\frac
12)\sin^2\theta\cos^2\phi + \frac j2,\nonumber\\
N^2 \eval{z}{\hat J_2^2}{z}&=& j(j-\frac
12)\sin^2\theta\sin^2\phi + \frac j2.
\eea
Thus $N^2 \eval{z}{\hat {\bf J}^2}{z}=j(j+1)$, as it should.

The normalized wavefunctions
$ 
N(\z,z)\psi^{(1)}_m(z)
$
have their maximum amplitude on the  lines of latitude
\be
\vert z\vert^2 =\vert z_m\vert^2 = \frac{j+m}{j-m}
\ee
corresponding to the polar angle $\theta_m =\cos^{-1}m/j$.
Note that 
\be 
N^2 \eval{z_m}{\hat J_3}{z_m}
=j \frac {\vert z_m\vert^2 -1}{\vert z_m\vert^2 +1}=m.
\ee
The variance, in terms of $m$, is given by  
\be
\left(N^2\vev{\hat J_3^2}- N^4\vev{\hat J_3}^2\right)=
\frac 12 j\left(1-\left(\frac {\z z-1}{\z z+1}\right)^2\right)
=\frac 12 j (1-\cos^2\theta).
\ee
Since  $m\sim j\cos \theta$, 
the normalized wavefunctions   
have zonal  spread $\Delta \theta \sim 1/\sqrt{j}$.
As $j$ becomes large the quantum spin becomes more
localized, and more
classical.

\section{Spin Action}

We wish to find a  semiclassical approximation  for the propagator
\be
K(\bzeta_f,\zeta_i,T)= \eval{\zeta_f}{e^{-i\hat H T}}{\zeta_i}
\label{EQ:propagator}
\ee
in the form
\be 
K_{\rm scl}(\bzeta_f,\zeta_i,T)= K_{\rm reduced}\cdot\exp\left\{S_{\rm
cl}(\bzeta_f,\zeta_i,T)\right\}.
\ee
Here $S_{\rm cl}$ is the   action for a classical path
going  from the point $z=\zeta_i$ to  the point $z=\zeta_f$
in time $T$. The action functional is expected to be that 
appearing in the  path integral representation of the 
exact propagator. The  amplitude $K_{\rm reduced}$,  the
{\it pre-exponential factor\/}, is then given by  a
gaussian approximation to  the integral over  deviations
from  the classical trajectory.  Such a    semiclassical   
approximation should be  accurate when $j$ is 
large.

If a continuous-time  path integral is ``derived'' by
inserting  $N$ intermediate overcompleteness relations into
(\ref{EQ:propagator}) and taking a
formal limit $N\to \infty$, then we find\cite{kochetov95}
\be
K(\bzeta_f,\zeta_i,T)=\int_{\zeta_i}^{\bzeta_f}d\mu(\z,z)
\exp\{S(\z(t), z(t))\},
\ee
where the path measure $d\mu$ is 
\be
d\mu(\z(t),z(t))=\lim_{N\to\infty} \prod_{n=1}^N\frac{2j+1}{\pi}
\frac{d^2z_n}{(1+\z_n z_n)^{2j+2}},
\ee
and the 
action $S(\z(t),z(t))$ is given by  
\be
S(\z(t),z(t)) = j\left\{\ln(1+ \bzeta_f z(T)) 
+\ln(1+ \z(0)\zeta_i)\right\}
+
\int_0^T \left\{ j \frac{\dot\z z- \z \dot z}{1+\z z} 
-i H(\z,z)\right\} dt.
\ee
Here the classical hamiltonian, $H(\z,z)$, is related to the
quantum $\hH$ by 
\be
H(\z,z)= \eval{z}{\hH}{z}/\brak{z}{z}.
\ee
The paths $z(t)$, $\z(t)$ obey the boundary conditions
$z(0)=\zeta_i$, $\z(T)=\bzeta_f$, but $\z(0)$, $z(T)$, being
actually $\z(0+\epsilon)$ and $z(T-\epsilon) $, are unconstrained, 
and are to be integrated over\cite{kochetov95}.  

When we regard  $S$ as the phase-space action for a 
classical system\cite{balachandran83},  the explicit
boundary terms, which appear naturally in the discretized
path integral, serve to ensure that both the first-order
Hamilton equations and their boundary conditions are
compatible with the action principle.  To see this, make  a
general variation in   the  trajectory, including
variations in the endpoints. We find that
\bea 
\delta S &=&
\frac{2j z(T)}{1+ \bzeta_f z(T)}\delta \bzeta_f + 
\frac{2j \z(0)}{1+ \z(0)\zeta_i} \delta \zeta_i\nonumber\\
&&+ 
\int_0^T \left\{ 
\delta z(t)\left( \frac {2j\dot\z}{(1+\z z)^2}-
i \frac{\partial H}{\partial z}\right)
+ 
\delta \z(t)\left( -\frac {2j\dot z}{(1+\z z)^2}-
i \frac{\partial H}{\partial \z}\right)\right\}dt.
\label{EQ:hamilton-jacobi}
\eea
There are no boundary contributions proportional to
$\delta\z(0)$ or $\delta z(T)$ because of a cancellation of
such terms arising from an integration by parts against
those arising from the variation of the explicit boundary
terms. Equating the variation of the action to zero
therefore requires  the classical path to obey  the
Hamilton equations
\be
\dot\z =i \frac {(1+\z z)^2}{2j}\frac{\partial H}{\partial z},
\quad 
\dot z =- i \frac {(1+\z z)^2}{2j}\frac{\partial H}{\partial \z},
\ee
together with  boundary conditions that fix  $z(0)=\zeta_i$, and 
$\z(T)=\bzeta_f$.
 
The quantities $\z(0)$ and $z(T)$ are not 
fixed by the boundary conditions, but can be found by
solving the equations of motion. If we know the action for
the classical path, they can also be read off from  the
Hamilton-Jacobi  equations that follow from  
(\ref{EQ:hamilton-jacobi}), {\it viz:\/}  
\be
\frac{\partial S_{\rm cl}}{\partial \bzeta_f}= \frac {2jz(T)}{1+ \bzeta_f z(T)},
\quad
\frac{\partial S_{\rm cl}}{\partial \zeta_i}= \frac {2j\z(0)}{1+
\z(0)\zeta_i}.
\label{EQ:jacobi_equations}
\ee
In general $\z(0)$ will not be the complex conjugate of
$z(0)\equiv\zeta_i$, nor will $z(T)$ be the complex conjugate of
$\z(T)\equiv \bzeta_f$. This means that if we write  $z$ as
$x+iy$ and $\z=x-iy$, then, except in special cases, $x$
and $y$ are not real numbers.

The Hamilton-Jacobi relations also tell us that  
\be 
\frac{\partial S_{\rm cl}}{\partial \bzeta_i}=
\frac{\partial S_{\rm cl}}{\partial
\zeta_f}=0,
\ee
showing that $S_{\rm cl}$ is a holomorphic function of
$\zeta_i$, and an anti-holomorphic function of $\zeta_f$.
These  analyticity properties of $S_{\rm cl}$ coincide with
those of $K$. This is reasonable since $\exp S_{\rm cl}$ 
is the leading approximation to $K$, and we would expect
analyticity to preserved term-by-term in the large $j$
expansion. Finally 
\be  
\frac{\partial S_{\rm cl}}{\partial T}= -i H(\bzeta_f,
z(T)).
\ee

The 
leading semiclassical approximation is
exact when the quantum hamiltonian $\hH$ is an element of the Lie algebra of $SU(2)$.
For example, 
if $\hat H= \omega \hat J_3$, then
\be
H(\z,z)=N^2\eval{z}{\hat H}{z} = \omega j \frac{\z z -1}{\z z+1}
\ee
and 
\be 
\frac{\partial H}{\partial z}= \frac {2j\omega\z}{(1+ \z
z)^2},\qquad \frac{\partial H}{\partial \z}= \frac {2j\omega
z}{(1+ \z z)^2},
\ee
The equations of motion are therefore 
\be
\dot\z =i\omega \z, \quad \dot z = -i\omega z.
\ee
The solutions  obeying the appropriate boundary conditions are
\be
z(t) = e^{-i\omega t}\zeta_i, \quad \z(t) = e^{i\omega(t-T)}\bzeta_f,
\label{EQ:classical_solns}
\ee
so
\be
z(T) = e^{-i\omega T}\zeta_i, \quad \z(0) =  e^{-i\omega T}\bzeta_f.
\ee
It will only be in exceptional circumstances that $z(T)= (\bzeta_f)^*$ or
$\z(0)= (\zeta_i)^*$. 

Inserting the solutions (\ref{EQ:classical_solns}) into the action  we find 
\bea
S_{\rm cl}(\bzeta_f,\zeta_i, T)&=&
j\left\{\ln(1+ \bzeta_f \zeta_i e^{-i\omega T} ) 
+\ln(1+ \bzeta_f \zeta_i e^{-i\omega T} ) \right\}
+
\int_0^T \left\{ i j\omega  \frac {2\z z}{1+\z z } -ij \omega \frac{\z z -1}{
z z +1}  \right\} dt\nonumber\\
&=& 2j\ln(1+ \bzeta_f \zeta_i e^{-i\omega T}) + i j\omega T.
\eea
This is to be compared with the  exact propagator  
\be
K=\eval{\zeta_f}{e^{-i\hat H T}}{\zeta_i}= e^{i\omega j T
}(1+e^{-i\omega T} \bzeta_f \zeta_i)^{2j} =\exp S_{\rm cl}.
\ee

When the hamiltonian is a more  general element of the enveloping
algebra ({\it i.e.\/} a polynomial in the generators) there will be
corrections to this simple result.

\section{Fluctuation Determinant}

The prefactor in the semiclassical propagator comes from
integration over  gaussian fluctuations about the classical
trajectory. To evaluate these, we consider the  second
variation of the classical action, holding $z(0)=\zeta_i$
and $\z(T)=\bzeta_f$  fixed. 
We will  write 
\be
S=S_{\rm cl}+ \delta S +\frac 12 \delta^2
S+\cdots,
\ee
where 
\be
\delta^2 S= -i\int_0^T \frac {2j}{(1+\z z)^2}  \left(\matrix{ \delta\z &  \delta z\cr}\right)
\left[\matrix{-i\partial_t + A & B \cr
                \B & i\partial_t +A \cr}\right]
\left(\matrix{ \delta z \cr \delta \z}\right)dt.
\label{EQ:delta2S}
\ee
Here
\bea
A &=& \frac 12 \left(\frac{\partial}{\partial \z} \frac{(1+ \z z)^2}{2j}
\frac{\partial H}{\partial z} + \frac{\partial}{\partial z} \frac{(1+ \z z)^2}{2j}
\frac{\partial H}{\partial \z}\right), \nonumber\\
B &=& \frac{\partial}{\partial \z} \frac{(1+ \z z)^2}{2j}
\frac{\partial H}{\partial \z}, \nonumber\\
\B &=& \frac{\partial}{\partial z} \frac{(1+ \z z)^2}{2j}
\frac{\partial H}{\partial z}.
\label{EQ:D_coefficients}
\eea
When   $z(t)$, $\z(t)$ are   the classical path, then 
$\delta S=0$.

On making  a  change of variables
\bea
\delta z &=& (1+\z z) \eta,\nonumber\\
\delta \z &=& (1+\z z) \bar\eta,
\eea 
we see that we have to compute the quadratic path integral
\be
K_{\rm reduced}\propto \int d[\eta]d[\bar\eta] \exp 
 \left\{-2ij \int_0^T  \frac 12 \left(\matrix{ \bar\eta &  \eta\cr}\right)
\left[\matrix{-i\partial_t+ A & B \cr
              \B & i\partial_t+A \cr}\right]
\left(\matrix{ \eta \cr \bar\eta}\right)dt\right\}.
\label{EQ:eta_path_integral}
\ee
This path integral is  proportional to $\Det^{-\frac 12} {\cal D}$, where 
the matrix differential operator 
\be
{\cal D}= \left[\matrix{-i\partial_t+ A & B \cr
              \B & i\partial_t+A \cr}\right] = -i\sigma_3
	      \partial_t +M 
\ee
is subject to  the boundary conditions $\eta(0)=0$ and $\bar\eta(T)=0$. 
(We will use the symbol ``$\Det$'' for functional determinants and 
``$\det$''
to denote  the determinant of a finite matrix. Similarly ``$\Tr$'' and 
``$\tr$''.)

There are several subtleties involved in calculating $\Det
{\cal D}$.   The most obvious is that  the boundary
conditions imposed on ${\cal D}$ are {\it not} in the class
that make it self adjoint. Although ${\cal D}$ and ${\cal
D}^\dagger $ are formally the same differential operator, 
self-adjointness requires, in addition, that their domains
of definition coincide\cite{richtmeyer78}. It is not
hard to see that the only  boundary
condition on ${\cal D}$ that leads to  an identical boundary
condition for ${\cal D}^\dagger$ is  $\eta(0)=
e^{i\theta_0} \bar\eta(0)$ and  $\eta(T)= e^{i\theta_T}
\bar\eta(T)$ for some real angles $\theta_0$, $\theta_T$.
Indeed if $B=\B=0$, then  ${\cal D}$ with our boundary
conditions  has {\it no\/} eigenfunctions --- never mind a
complete set.  The determinant cannot be expressed as  an
infinite product of eigenvalues, therefore. 
Diagonalizability is not, however,  a fundamental
requirement for defining a determinant. There exists a
well-defined Green function $G={\cal D}^{-1}$, and we
should be able to  obtain the determinant by varying the
parameters and  using the identity $\delta \ln \Det {\cal
D} = \Tr \{ {\cal D}^{-1} \delta{\cal D}\}$, which holds
even if $\cal D$ is  not diagonalizable.  

A potential   pitfall in this approach is that the variation 
$\delta
\ln \Det {\cal D} $ is given by 
\be
\delta\ln \Det {\cal D} =\Tr \{ {\cal D}^{-1} \delta{\cal D}\}= 
\int_0^T \tr\{ G(t,t)\delta M\}dt,
\ee
but the Green function $G(t,t')$ is discontinuous at
$t=t'$. We might have a different expression for the
variation  depending on whether we choose to evaluate
$G(t,t)$ as $G(t,t+\epsilon)$ or as $G(t,t-\epsilon)$. The
jump in $G$ is, however, proportional to $\sigma_3$, and
$\tr\{\sigma_3 \delta M\}\equiv 0$, so we have reason to
hope that  there is no actual ambiguity.   

If we agree to interpret $G(t,t)$ as  $\frac 12
(G(t,t+\epsilon)+G(t,t-\epsilon))$, then the  formal
calculation is  straight forward\cite{stone-kos98},  and we
merely summarize the results:

We begin by defining  the matrix  
\be 
\Phi(t)=\left(\matrix{ \eta_T(t) & \eta_0(t)\cr
                    \bar\eta_T(t) & \bar\eta_0(t)\cr}\right).
\ee   
Here the column vector $(\eta_0(t),\,\bar\eta_0(t))^T$ is a solution 
of ${\cal D}\Psi=0$
obeying the boundary condition $\eta_0(0)=0$, $\bar\eta_0(0)=1$, and
$(\eta_T(t),\,\bar\eta_T(t))^T$ is a solution with $\eta_T(T)=1$, 
$\bar\eta_T(T)=0$. The determinant of  $\Phi(t)$ is an analogue of the 
wronskian and  is  independent of $t$. We find that  
$\Det {\cal D}= C \det \Phi$, where $C$ is some constant
independent of $H$.

Since $\det\Phi$ is time independent, we may conveniently evaluate it at
$t=T$, where 
\be
C^{-1} \Det {\cal D}=   \left |\matrix{ 1 & \eta_0(T)\cr
                                  0 & \bar \eta_0(T)\cr}\right|= 
\bar\eta_0(T),
\ee
or at $t=0$, where   
\be
C^{-1} \Det {\cal D}=   \left |\matrix{ \eta_T(0) & 0\cr
                              \bar\eta_T(0) & 1\cr}\right|=
\eta_T(0).
\ee

By  relaxing the conditions that $\eta(T)=\bar\eta(0)=1$, 
we may interpret these results in terms of the  
variation of the endpoints of the classical trajectory as we vary the 
initial points. That is   
\be
C^{-1} \Det {\cal D}= \left(\frac{\partial\bar\eta(0)}{\partial
\bar\eta(T)}\right)^{-1} = \left(\frac{\partial\eta(T)}{\partial
\eta(0)}\right)^{-1},
\ee
or, in terms of the original variables,
\be 
C^{-1} \Det {\cal D}= 
\frac{1+\z(0) \zeta_i}{1+ \bzeta_f z(T)}
\left(\frac {\partial\z(0)} {\partial\bzeta_f}\right)^{-1}= 
\frac{1+ \bzeta_f z(T)}{1+ \z(0)\zeta_i }
\left(\frac{\partial z(T)} {\partial\zeta_i}\right)^{-1}.
\label{EQ:DetD}
\ee

The equivalence of these two expressions for the determinant is not
immediately obvious, but 
from the Hamilton-Jacobi relations
\be
\frac{\partial S_{\rm cl}}{\partial \bzeta_f}= 
\frac {2jz(T)}{1+ \bzeta_f z(T)},
\quad
\frac{\partial S_{\rm cl}}{\partial \zeta_i}= \frac {2j\z(0)}{1+
\z(0)\zeta_i},
\ee 
and the equality of mixed partials of $S_{\rm cl}$, we obtain 
\be
\frac {\partial^2 S_{\rm cl}}{\partial \zeta_i\partial \bzeta_f}=
\frac {2j}{(1+\bzeta_f z(T))^2}\frac{\partial z(T)} {\partial\zeta_i}
= \frac{2j}{(1+\z(0)\zeta_i)^2 }\frac {\partial \z(0)}
{\partial\bzeta_f}.
\ee
Both expressions in (\ref{EQ:DetD}) thus reduce to  
\be
C \Det^{-1} {\cal D}= \frac{
(1+\bzeta_f z(T))(1+\z(0)\zeta_i)}{2j}
\frac {\partial^2 S}{\partial \zeta_i\partial \bzeta_f}.
\ee

Our calculation of the fluctuation determinant  suggests, therefore, that
\be 
K_{\rm scl}(\bzeta_f,\zeta_i,T)\stackrel{?}{=}
\left(\frac{(1+\bzeta_f z(T))(1+\z(0)\zeta_i)}{2j}
\frac {\partial^2 S_{\rm cl}}{\partial \zeta_i\partial
\bzeta_f}\right)^{\frac 12}
\exp {S_{\rm cl}(\bzeta_f, \zeta_i, T)}.
\label{EQ:naive_prop}
\ee
(The proportionality constant  is fixed by the requirement 
that this expression reduces 
to $\brak{\zeta_f}{\zeta_i}$ when $T=0$).

As indicated  by the ``?'' over the equals sign, there are  problems with this expression, 
and it is not quite correct.  
 
The first problem is that, despite the  optimism expressed
above, there {\it is} a degree of indeterminacy in the
calculation of the functional  determinant.  To see this,
make the substitution
\bea
\eta(t) &\to& e^{i\theta(t)}\eta(t),\nonumber\\
\bar\eta(t) &\to& e^{-i\theta(t)}\bar\eta(t).
\label{EQ:eta_to_etaprime}
\eea 
in the path integral (\ref{EQ:eta_path_integral}).
The measure is unchanged,
but we replace  ${\cal D}$ with  $\tilde{\cal D}$, 
where $\tilde{\cal D}$ is the
matrix operator $\cal D$ with  
\bea
A\to \tilde A &=& A + \partial_t\theta, \nonumber\\
B\to \tilde B &=& e^{-2i\theta(t)}B,\nonumber\\
\B\to \tilde \B &=& e^{2i\theta(t)} \B.
\label{EQ:D_toDprime}
\eea
The value of the path integral must be  unaltered by this
change of integration variables, but    
the solution to 
\be 
 \left[\matrix{-i\partial_t+ \tilde A & \tilde B \cr
              \tilde \B & i\partial_t+\tilde A \cr}\right]
\left(\matrix{\eta(t)\cr \bar \eta(t)}\right)=0
\ee
with $\eta(0)=0$, $\bar\eta(0)=1$ is now
$(e^{-i(\theta(t)-\theta(0))}\eta_0(t), 
e^{i(\theta(t)-\theta(0))}\bar\eta_0(t))^T$. The determinant,
as we have calculated it, 
is therefore {\it not\/}
invariant, but ends up multiplied 
by $e^{-i(\theta(T)-\theta(0))}$. Our expression for the  functional   
determinant   has  an 
``anomaly'' therefore. 

The anomaly arises because the argument we made about the
harmlessness of the discontinuity in $G$ depends on our
defining  $G(t,t)$ as $G(t,t\pm \epsilon)$ with  the {\it
same choice of sign\/} in front of the  $\epsilon$ in  both
entries in the trace. If we examine the  discrete version
of path integral we see that, on the contrary, one of the
entries should be  evaluated with a plus, and one with a
minus. Our calculation  of the determinant assumed that we
could  interpret $G(t,t)$ as $\frac 12(G(t,t+\epsilon)+
G(t,t-\epsilon))$, so our formula for the determinant is 
only correct if both terms in $\tr\{\sigma_3 \delta M\}$
are separately zero. This will only be the case   for 
operators ${\cal D}$ with $A\equiv 0$. Fortunately the
discrete path integral {\it does} permit the change of
variables described above, and we may  use  this freedom to
force
the diagonal entries, $\tilde A$, to zero before computing
the determinant. The correctly regulated  functional 
determinant therefore differs from  its na{\"\i}ve value by
a multiplicative factor.       

Including the correction to the  fluctuation determinant, the 
semiclassical propagator  becomes  
\be
K_{\rm scl} (\bzeta_f,\zeta_i,T)=
\left(\frac{(1+\bzeta_f z(T))(1+\z(0)\zeta_i)}{2j}
\frac {\partial^2 S_{\rm cl}}{\partial \zeta_i\partial
\bzeta_f}\right)^{\frac 12} \exp\left\{S_{\rm cl}(\bzeta_f, \zeta_i, T)
+\frac{i}{2}\int_0^T A(t)dt\right\}.
\label{EQ:kochetov_propagator}
\ee
where  
\be 
A(\z,z)=  \frac 12 \left(\frac{\partial}{\partial \z} \frac{(1+ \z z)^2}{2j}
\frac{\partial H}{\partial z} + \frac{\partial}{\partial z} \frac{(1+ \z z)^2}{2j}
\frac{\partial H}{\partial \z}\right),
\ee
is the coefficient  appearing in (\ref{EQ:D_coefficients})

The manoeuvre of setting $\tilde A$ to zero before
evaluating the fluctuation determinant appears   (although
without explanation as to why it was necessary) in the
previously cited  paper
by   Kochetov\cite{kochetov95}  that provided part of the
motivation for our present work. Kochetov  therefore gets
the corrected expression (\ref{EQ:kochetov_propagator}). It
seems, however, that   the ``extra phase'' (it is a phase
only in the simplest cases), $\frac{i}{2}\int_0^T A(t)dt$,
was first obtained by Solari\cite{solari87} from a careful
evaluation  of the discrete determinant. Solari also
pointed out the necessity  of  a similar correction in the
harmonic oscillator coherent-state path integral, which has
a flat phase space.   Kochetov's discovery of the
correction seems to have been independent of this earlier work.

Because of the extra phase,  (\ref{EQ:kochetov_propagator})
gives the correct,  indeed exact, semiclassical propagator
for the case $\hH=\omega \hat J_z$, and also for any
hamiltonian consisting of (possibly time dependent)
elements of the Lie algebra of $SU(2)$ \cite{kochetov95}.

\section{Short Time Accuracy}

The Solari-Kochetov phase  also solves a second problem
with  (\ref{EQ:naive_prop}). In contrast to the
configuration space propagator, which diverges  as
$T^{-\frac 12}$, the coherent-state propagator
$K(\zeta_f,\zeta_i,T)$ is analytic in $T$ near $T=0$. This
is because of the finite spread of the coherent-state
wavefunctions. To first order in $T$ we have
\bea
K(\bzeta_f,\zeta_i,T)\equiv \eval{\zeta_f}{e^{-i\hat H T}}{\zeta_i}
&\approx& \brak{\zeta_f}{\zeta_i} -iT\eval{\zeta_f}{\hH}{\zeta_i}\nonumber\\ 
&=&
\brak{\zeta_f}{\zeta_i}\left(1-iTH(\bzeta_f,\zeta_i)\right).
\label{EQ:short_time_limit}
\eea
(In the last equality we have exploited  analyticity to observe that 
the off-diagonal   $\eval{\zeta_f}{\hH}{\zeta_i}$,
is obtained from the diagonal 
$\eval{\zeta}{\hH}{\zeta}$ by the  the simple replacement $\zeta \to \zeta_i$,
$\bzeta \to \bzeta_f$ .)

Now, from the Hamilton-Jacobi equation
\be  
\frac{\partial S_{\rm cl}}{\partial T}= -i H(\bzeta_f,
z(T)),
\ee
we have 
\be
S_{\rm cl}(\bzeta_f, \zeta_i, T) =
S_{\rm cl}(\bzeta_f, \zeta_i, 0)-iT H(\bzeta_f,\zeta_i)+O(T^2),
\ee 
while
\be
S_{\rm cl}(\bzeta_f, \zeta_i, 0)=2j\ln(1+\bzeta_f\zeta_i)=
\ln\brak{\zeta_f}{\zeta_i}.
\ee 
Thus, in order to get agreement between
(\ref{EQ:kochetov_propagator}) and  (\ref{EQ:short_time_limit}), the
fluctuation determinant must make no $O(T)$ contribution to the propagator.
A short calculation shows, however, that
\be
 \frac{(1+\bzeta_f z(T))(1+\z(0)\zeta_i)}{2j}
\frac {\partial^2 S_{\rm cl}}{\partial \zeta_i\partial \bzeta_f}
= 1-iT A(\bzeta_f,\zeta_i) +O(T^2).
\ee
Fortunately this contribution is exactly cancelled by the
$O(T)$ contribution from the Solari-Kochetov extra phase.

We now  ask how well does  the  semiclassical propagator
do at next order in the short-time expansion. 
In order to  provide a systematic grading for  the terms,
we will regard  the hamiltonian $\hH$ as being $O(j)$. The
entire action is then  homogeneous of degree one in $j$. 
With this assumption, and by analogy with the usual
semiclassical expansion in powers of $\hbar$,  we  expect
that
\be 
K(\bzeta_f, \zeta_i,T) = K_{\rm reduced}\cdot\exp\{S_{\rm
cl}\}\cdot
\left[1+O\left(\frac 1j\right)\right],
\ee 
where $S_{\rm cl}$  is   $O(j)$, while the prefactor, $K_{\rm
reduced}$, is $O(j^0)$.

At short time the exact coherent-state propagator is certainly 
of this
form. To demonstrate this, expand   
\be
\eval{\zeta_f}{e^{-i\hH T}}{\zeta_i}=
\brak{\zeta_f}{\zeta_i}-iT\eval{\zeta_f}{\hH
}{\zeta_i}-\frac{T^2}{2}
\eval{\zeta_f}{\hH^2}{\zeta_i}+\cdots.
\ee
Now $
\eval{\zeta_f}{\hH}{\zeta_i}=
\brak{\zeta_f}{\zeta_i}H(\bzeta_f,\zeta_i)
$,
but some work is needed to  evaluate $\eval{\zeta_f}{\hH^2}{\zeta_i}$.

Inserting an overcompleteness integral, we have
\bea
\eval{\zeta_f}{\hH^2}{\zeta_i}&=&
\frac{2j+1}{\pi} \int \frac {d^2z} {(1+\z
z)^{2j+2}}\eval{\zeta_f}{\hH}{z}\eval{z}{\hH}{\zeta_i}
\nonumber\\
&=& 
\frac{2j+1}{\pi} \int \frac {d^2z} {(1+\z
z)^{2j+2}}(1+\bzeta_f z)^{2j}(1+\z
\zeta_i)^{2j}H(\bzeta_f,z)H(\z,\zeta_i).
\eea
We now perform a steepest descent expansion  in the 
integral over the intermediate states, and obtain the 
first three terms in its  asymptotic expansion in powers of
$1/j$. This computation is greatly simplified by using  two
shortcuts: First  we need calculate only the diagonal
matrix element $\eval{\zeta}{\hH^2}{\zeta}$. Given this, we
may  appeal to analyticity  and obtain the general matrix
element by setting  $\bzeta \to \bzeta_f$ and $\zeta\to
\zeta_i$. Next we rotate the sphere so as to centre  the
coordinate system on the point $\zeta$. Thus  
$\zeta\to 0$, and the coordinate system is locally geodetic. 
In these coordinates  the saddle point of the
$z$ integral is at $\zeta=z=0$, and far fewer
terms have to taken into consideration.

To return to the original coordinates, we need to
be able to recognize some $SO(3) \simeq SU(2)$ invariant 
combinations of  derivatives and $(1+\z z)^2$ factors.

One easily establishes that, under the M\"obius mapping
\be
z\to z' = \frac{az+b}{cz+d}, 
\quad \hbox{\rm where}\quad \left[\matrix{a&b\cr
c&d\cr}\right] \in SU(2), 
\ee
we have
\be
\frac{d^2z}{(1+\z z )^2}= \frac{d^2z'}{(1+\z' z'
)^2}
\ee
together with 
\bea
(1+\z z)^2 \frac{\partial f(\z, z )}{\partial z }
\frac{\partial g(\z, z )}{\partial \z}&=&
(1+\z' z')^2 \frac{\partial f(\z', z' )}{\partial z' }
\frac{\partial g(\z', z' )}{\partial \z' },\nonumber\\
(1+\z z)^2 \frac{\partial^2 f(\z, z )}{\partial z\partial \z }
&=&(1+\z' z')^2 \frac{\partial^2 f(\z,' z' )}{\partial
z'\partial \z' },
\eea
and that the combination
\be
Z= \left(\frac{\partial }{\partial z }(1+\z z)^2\frac{\partial
 }{\partial z } f\right)\left(\frac{\partial }{\partial \z }(1+\z z)^2\frac{\partial
 }{\partial \z } g\right)
 \label{EQ:invcom}
 \ee
is similarly invariant. Thus, when we see 
the term $\partial^2_{zz}f\partial^2_{\z \z} g$ appearing in the
expansion about the stationary point $z=0$, we realize 
that in the integral for  the general matrix element 
(where the saddle point is at
$z=\zeta_i$, $\z=\bzeta_f$) we
should replace it by (\ref{EQ:invcom}). 

Proceeding in this manner we find
\bea
\eval{\zeta_f}{\hH^2}{\zeta_i}&=&\quad \brak{\zeta_f}{\zeta_i}
\left\{ H^2(\bzeta_f,\zeta_i)+\frac{(1+\bzeta_f\zeta_i)^2}{2j} \frac{\partial
H}{\partial\zeta_i}\frac{\partial
H}{\partial\bzeta_f}\right.\nonumber\\
&&\qquad \left. + \frac 12 \frac{1}{(2j)^2} 
\left(\frac{\partial}{\partial\bzeta_f}(1+\bzeta_f\zeta_i)^2
\frac{\partial H}{\partial\bzeta_f}\right)
\left(\frac{\partial}{\partial\zeta_i}(1+\bzeta_f\zeta_i)^2
\frac{\partial H}{\partial\zeta_i}\right) + O\left(\frac
1j\right)\right\}.
\label{EQ:matrixHsq}
\eea 
The three terms in braces in this expression are of
$O(j^2)$, $O(j)$, and of $O(j^0)$ respectively.

We may now re-exponentiate (\ref{EQ:matrixHsq}) as
\bea
\eval{\zeta_f}{e^{-i\hH T}}{\zeta_i}&=&
\exp\left\{\ln \brak{\zeta_f}{\zeta_i} -iT
H(\bzeta_f,\zeta_i)
-\frac 12 T^2 \frac{(1+\bzeta_f\zeta_i)^2}{2j} \frac{\partial H
}{\partial \zeta_i }\frac{\partial H }
{\partial \bzeta_f }+\cdots\right\}\nonumber\\
&&\times\left[ 1-\frac {T^2}{4}\cdot \frac 1{(2j)^2}
\left(\frac{\partial }{\partial \zeta_i }(1+\bzeta_f \zeta_i)^2\frac{\partial
}{\partial \zeta_i } H\right)\left(\frac{\partial }{\partial
\bzeta_f }(1+\bzeta_f \zeta_i)^2\frac{\partial
}{\partial \bzeta_f } H\right)+\cdots\right]
\label{EQ:expform} 
 \eea 

Again using the Hamilton Jacobi equation,
\be  
\frac{\partial S_{\rm cl}}{\partial T}= -i H(\bzeta_f,
z(T)),
\ee
and  the equation of motion for $z(t)$, we may  generate the Taylor
series for $S_{\rm cl}(T)$.
We immediately verify the term in the exponential is the
classical action to $O(T^2)$:
\be
S_{\rm cl}=\ln \brak{\zeta_f}{\zeta_i} -iT
H(\bzeta_f,\zeta_i)
-\frac 12 T^2 \frac{(1+\bzeta_f\zeta_i)^2}{2j} 
\frac{\partial H }{\partial \zeta_i }\frac{\partial H }
{\partial \bzeta_f }+O(T^3).
\ee
The expression  in the square brackets in (\ref{EQ:expform})
must be  the prefactor,
and is manifestly  $O(j^0)$. 
It is a little tedious to verify that our
formula for the pre-exponential factor, including the Solari-Kochetov
correction, reduces to exactly this, but it is so. To
collapse the terms, it helps to 
use the identity
\bea
&&(1+\z z)^2 \frac{\partial^2}{\partial\z\partial z} 
\left((1+\z z)^2 \frac{\partial H}{\partial\z}\frac{\partial
H}{\partial z}\right)= \nonumber\\
&&\qquad 2(1+\z z)^2\frac{\partial H}{\partial\z}\frac{\partial
H}{\partial z}+ \left((1+\z z)^2 \frac{\partial^2H}{\partial\z\partial
z}\right)^2
+(1+\z z)^2 \frac{\partial H}{\partial\z}\frac{\partial}{\partial z}
\left( (1+\z z)^2 \frac{\partial^2H}{\partial\z\partial
z}\right)\nonumber\\
&&\qquad +(1+\z z)^2 \frac{\partial H}{\partial z}\frac{\partial}{\partial\z}
\left( (1+\z z)^2 \frac{\partial^2H}{\partial\z\partial
z}\right)+\left(\frac{\partial }{\partial z }(1+\z z)^2\frac{\partial
 }{\partial z } H\right)\left(\frac{\partial }{\partial \z }(1+\z z)^2\frac{\partial
 }{\partial \z } H\right),
\eea
which is most easily established by noting that all terms
are invariant, and, at $z=0$, both sides reduce to
\be
\left(\frac{\partial^2}{\partial\z\partial
z}+2 \right)\frac{\partial H}{\partial\z}\frac{\partial
H}{\partial z}.
\ee 

The  semiclassical expression, therefore,  has errors of at most 
$O(j^{-1})$
at short time. Our expectation  is, of course, that it has
this degree of accuracy uniformly in $T$.

\section{Consistency}

A further  test of the correctness of (\ref{EQ:kochetov_propagator}) is to
verify its consistency under dissection of the classical trajectory.
The exact propagator must satisfy the sewing condition 
\be
K(\bzeta_f,\zeta_i,t_1+t_2)= \frac {2j+1}{\pi} \int
\frac{d^2\xi}{(1+\bxi \xi)^{2j+2}}
K(\bzeta_f,\xi,t_2)K(\bxi,\zeta_i,t_1), 
     \label{stitch}
\ee 
which follows from the definition of $K$ and the
overcompleteness condition (\ref{EQ:overcompleteness}). 
The  semiclassical approximation to $K$ should obey a
similar condition, but with the exact integration over the
intermediate states replaced by a suitable stationary phase
approximation. 

Since  $K_{\rm scl}\sim
\exp S_{\rm cl}$, we  begin with the relationship  between the
action for the total path from $\zeta_i$ to $\zeta_f$, and the
actions for the two segments from $\zeta_i$ to the intermediate point
$\xi$, and from $\xi$ to $\zeta_f$. To eliminate the redundant
intermediate-point  boundary terms we must define 
\be
S(\bzeta_f,\zeta_i, t_1+t_2)= S(\bzeta_f, \xi,t_2)+
S(\bxi,\zeta_i,t_1) -2j\ln(1+\bxi \xi).  
\label{EQ:S_tot} 
\ee 
We will
write this compactly as 
\be 
S_{\rm tot}=S_2+S_1-2j\ln(1+\bxi \xi).
\ee

In writing  (\ref{EQ:S_tot}) we have tacitly assumed that our chosen
starting $\xi$ of the second path segment coincides with the
dynamically determined endpoint $z(t_1)$ of the first path segment,
and that the dynamically determined starting $\z(t_1)$ of the second
path segment coincides with our chosen $\bxi$ endpoint of the first
path segment. This will not generally be the case --- but it {\it
will\/} be when $\bxi$, $\xi$ obey the stationary-phase equations
\be
\frac{\partial S_{\rm tot}}{\partial \xi}= \frac{\partial S_{\rm tot}}{\partial
\bxi}=0.
\ee
Taking into account the analyticity properties of $S_1$ and $S_2$, these are  
\bea
0&=&\frac{\partial S_2(\bzeta_f,\xi)}{\partial \xi} - \frac{2j\bxi}{1+\bxi \xi},
\nonumber\\
0&=&\frac{\partial S_1(\bxi,\zeta_i)}{\partial \bxi} - \frac{2j\xi}{1+\bxi \xi}.    
\label{EQ:stat_phase}
\eea
Comparing (\ref{EQ:stat_phase}) with the Hamilton-Jacobi equations confirms 
that $\xi_c=z(t_1)$ and $\bxi_c=\z(t_1)$, where $\xi_c$, $\bxi_c$ is the 
stationary phase point.

To evaluate the integral over  small deviations from the classical stationary 
phase point, we set
$\xi=\xi_c+\eta$, $\bxi=\bxi_c+\bareta$.
We expand 
\be
S_{\rm tot}= S_{\rm tot}\vert_{\bxi_c,\xi_c} -\frac 12
\frac{2j}{(1+\bxi_c\xi_c)^2}
\left(\matrix{ \bareta, &\eta\cr}\right) 
\left[\matrix{ 1 & -\alpha\cr
               -\beta & 1 } \right]
\left(\matrix{\eta\cr
               \bareta\cr}\right),
\label{EQ:alpha_beta}
\ee
where
\be
\alpha= \frac{(1+\bxi_c\xi_c)^2}{2j}\frac{\partial^2 S_1}{\partial \bxi_c^2}+ 
\xi_c^2=\frac 1{2j}
(1+\bxi_c\xi_c)\frac{\partial }{\partial \bxi_c}(1+\bxi_c\xi_c)
\frac{\partial S_1}{\partial \bxi_c},
     \label{EQ:defalpha}
\ee
and
\be
\beta = \frac{(1+\bxi_c\xi_c)^2}{2j}\frac{\partial^2 S_2}{\partial \xi_c^2}+ 
\bxi_c^2=\frac 1{2j}
(1+\bxi_c\xi_c)\frac{\partial }{\partial \xi_c}(1+\bxi_c\xi_c)
\frac{\partial S_2}{\partial \xi_c}.
     \label{EQ:defbeta}
\ee
(The second equality in these equations uses the stationary phase equations.)

We now put together two semiclassical propagators
and perform the gaussian integral over the deviation from the stationary
phase point. Using the semiclassical Solari-Kochetov form  
(\ref{EQ:kochetov_propagator}) for the propagators on the right-hand
side of Eq.~(\ref{stitch}), we get (with $T = t_1 + t_2$),
\bea
K_{\rm comb} &=&\frac {2j+1}{\pi} \int \frac{d^2\eta}{(1+\bxi_c \xi_c)^2}
  \exp\left\{ S_1 + S_2 
-2j\ln(1+\bxi_c \xi_c)  +\frac{i}{2}\int_0^T A\,dt
-\frac 12 \delta^2 S \right\}\nonumber\\ 
&&
\times \left(\frac{(1+\bzeta_f z(T))(1+\bxi_c\xi_c)}{2j}
\frac{\partial^2 S_2}{\partial
\bzeta_f\partial\xi_c}
\frac{(1+\bxi_c\xi_c) (1+\z(0)\zeta_i)}{2j}
\frac{\partial^2  S_1}{\partial\bxi_c\partial\zeta_i}\right)^{\frac 12}. 
  \label{EQ:redivide-K}
\eea
Notice that, as with consistency test of the ordinary Feynman path
integral\cite{marinov80}, the measure and the prefactors,
including the Solari-Kochetov ``extra-phase'' term,  are
all being treated as constants. The integration involves
only the variation of the  classical
action
\be 
\delta^2  S=
\frac{2j}{(1+\bxi_c\xi_c)^2}
\left(\matrix{ \bareta, &\eta\cr}\right)
\left[\matrix{ 1 & - \alpha\cr
               - \beta & 1 } \right]
\left(\matrix{\eta\cr
               \bareta\cr}\right),
\ee                       
and yields, along with other factors, the inverse
square-root of the determinant
\be
D=\left |\matrix{ 1 & - \alpha\cr
                - \beta & 1 } \right|.
\ee	       

We now  use the result, established in the appendix, that
\be
\frac{\partial^2 S_{\rm tot}}{\partial \bzeta_f\partial\zeta_i}=
\frac{(1+\bxi_c\xi_c)^2}{2j}\frac{\partial^2 S_2}
{\partial \bzeta_f\partial\xi_c}
\frac{\partial^2 S_1}{\partial\bxi_c\partial\zeta_i}
\left\vert\matrix{ 1 & -\alpha\cr
               -\beta & 1 } \right\vert^{-1},
\label{EQ:KEY_EQ}
\ee
to obtain
\be
K_{\rm comb}=\left( \frac{2j+1}{2j}\right)
\left(\frac{(1+\bzeta_f z(T))(1+\z(0)\zeta_i)}{2j} \frac{\partial^2  S_{\rm tot}}{\partial \bzeta_f\partial\zeta_i}\right)^{\frac 12}
\exp\left\{ S_{\rm tot}(\bzeta_f,\zeta_i,T)+\frac i2\int_0^T
Adt\right\}.
\ee
The semiclassical approximation therefore reproduces itself 
except for a niggling
factor of $(2j+1)/2j$, which is due to a conflict between the normalization of
the measure and the $2j$ appearing in the exponent.

Although this discrepant factor approaches unity in the
large-$j$ limit, it is nonetheless disturbing. Each of the
infinitely many gaussian  integrations that constitute the
semiclassical approximation to the path integral  ought to  
be  indistinguishable from  our single gaussian integration
over the intermediate point $\xi$. We should, therefore, 
be able  to dissect the path into arbitrarily many parts
without affecting the final answer. This is not currently
so, and, in particular, the limit of large $j$ does not commute with the
limit of a large number of intermediate points.

The origin of the discrepancy is not hard to find. In the
large-$j$ limit
the effective radius of our spherical phase space 
becomes large, and, near $z=0$, the spin-$j$ reproducing-kernel relation
\be
\frac{2j+1}{\pi} \int \frac {d^2z}{(1+\z z)^2}
(1+\z z)^{-2j}
\brak{\zeta_2}{z}\brak {z}{\zeta_1} = \brak{\zeta_2}{\zeta_1},
\ee
or more explicitly,
\be
\frac{2j+1}{\pi} \int \frac {d^2z}{(1+\z z)^2} 
(1+\z z)^{-2j}(1+\bzeta_2 z)^{2j}(1+\z \zeta_1)^{2j}
=(1+\bzeta_2\zeta_1)^{2j},
\label{EQ:explicit}
\ee 
should contract to a suitably scaled version of its flat-phase-space analogue
\be
\int \frac {d^2z}{\pi} e^{-\z z} e^{\bzeta_2 z} e^{\z \zeta_1}
= e^{\bzeta_2\zeta_1}.
\label{EQ:flatrepro}
\ee
Because it is a gaussian integral, the leading stationary phase
``approximation'' to (\ref{EQ:flatrepro}) is exact.  

If we make the
obvious large $j$ estimates
\be
(1+\z z)^{-2j}\sim e^{-2j\z z},\quad (1+\bzeta_2 z)^{2j}\sim e^{2j\bzeta_2
z},\quad (1+\z \zeta_1)^{2j}\sim  e^{2j\z \zeta_1},
\ee
while regarding the sphere measure $(1+\z z)^{-2}$ as a prefactor, 
we do not get
exactly 
\be
\frac {2j}{\pi} \int d^2z \,e^{-2j\z z} e^{2j\bzeta_2 z} e^{2j\z \zeta_1}
= e^{2j\bzeta_2\zeta_1},
\ee
but instead $(2j+1)/2j$ times this.  

If we keep terms higher order in $1/2j$, both those  coming
from the measure and those from going beyond the
quadratic approximation to the exponent, they will of course
correct the error. What we really need, however,  
is a partitioning of the integral on the LHS of  (\ref{EQ:explicit}) 
such that the {\it leading\/}
steepest descent approximation will agree with the RHS. 
This will happen if  regard the expansion parameter as
$2j+1$ and not $2j$.
To see this, break up  
\be 
I =\frac{2j+1}{\pi} \int \frac {d^2z}{(1+\z z)^2} 
(1+\z z)^{-2j}(1+\bzeta_2 z)^{2j}(1+\z
\zeta_1)^{2j}
\ee
as
\be  
I= \frac{2j+1}{\pi} \int \frac {d^2z}{(1+\z z)^2} 
g^{-1}(\z, z) e^{(2j+1)\ln g(\z, z)}
\ee
with 
\be 
g(\z,z)= (1+\z z)^{-1}(1+\bzeta_2 z)(1+\z
\zeta_1).
\ee
The critical   point of the function in the
exponential is at $\z =\bzeta_2$, $z=\zeta_1$, and
\be 
g(\bzeta_2,\zeta_1)= (1+\bzeta_2\zeta_1),
\ee
\be
-\left.\frac{\partial^2 \ln g}{\partial z\partial \z}\right|_{\z
=\bzeta_2,z=\zeta_1}= \frac 1{(1+\bzeta_2\zeta_1)^2}.
\ee

Thus 
\bea
I &\sim& \frac{2j+1}{\pi} 
\frac {1}{(1+\bzeta_2\zeta_1)^3}(1+\bzeta_2\zeta_1)^{2j+1} 
\int \,d^2z 
e^{-\frac{2j+1}{(1+\bzeta_2\zeta_1)^2}(\z-\bzeta_2)(z-\zeta_1)} \nonumber\\
&=& \frac{2j+1}{\pi}
(1+\bzeta_2\zeta_1)^{2j-2}\cdot
\frac{\pi}{2j+1}(1+\bzeta_2\zeta_1)^2\nonumber\\   
&=& (1+\bzeta_2\zeta_1)^{2j}.
\label{EQ:2j+1decomposition}
\eea
The leading term of the asymptotic expansion of $I$ in powers of
$1/(2j+1)$ is therefore exact.

This observation  
suggests rewriting the   semiclassical approximation to $K$ as
\be
K_{\rm scl}(\bzeta_f,\zeta_i,T) =\frac 1{\sqrt{2j+1}}
\left(\frac{\partial^2 \tilde S_{\rm cl}}{\partial \bzeta_f\partial\zeta_i}\right)^{\frac 12}
\exp\left\{\tilde S_{\rm cl}(\bzeta_f,\zeta_i,T)+ \frac i
2\int_0^T Q dt\right\},
\label{EQ:consistentK}
\ee
where $\tilde S_{\rm cl}= (2j+1)S_{\rm cl}/(2j)$, and 
\be
Q=  \frac{1}{j}\left(
\frac {(1+\z z)^2}{2} \frac{\partial^2 H}{\partial \z\partial z} + H(\z ,z)\right) 
\ee
is the term required to make (\ref{EQ:consistentK}) numerically  equal to
(\ref{EQ:kochetov_propagator}). 

With this repartitioning of terms between the exponent and the
prefactor we have exactly the same classical equations of motion, but now 
\bea
K_{\rm comb} &=&\frac {2\jtil}{\pi} \int \frac{d^2\eta}{(1+\bxi_c \xi_c)^2} (1+\bxi_c
\xi_c) 
\exp\left\{\tilde S_1+\tilde S_2 -(2\jtil)\ln(1+\bxi_c \xi_c)
-\frac 12 \delta^2 \tilde S\right\}\nonumber\\ 
&& \times\frac 1{(2\jtil)}\left(\frac{\partial^2 \tilde S_2}{\partial \bzeta_f\partial\xi_c}
\frac{\partial^2 \tilde S_1}{\partial\bxi_c\partial\zeta_i}\right)^{\frac 12}
\exp\left\{\frac i2\int_0^T Qdt\right\}, 
  \label{EQ:Kcomb_j_tilde}
\eea
where
\be
  \jtil = j + \frac{1}{2},
\ee
and
\be 
\delta^2 \tilde S=
\frac{2\jtil}{(1+\bxi_c\xi_c)^2}
\left(\matrix{ \bareta, &\eta\cr}\right)
\left[\matrix{ 1 & -\tilde \alpha\cr
               -\tilde \beta & 1 } \right]
\left(\matrix{\eta\cr
               \bareta\cr}\right).
\ee
The quantities $\tilde\alpha$ and $\tilde\beta$ are obtained from
Eqs.~(\ref{EQ:defalpha}) and (\ref{EQ:defbeta})
by putting tildes on $S_1$, $S_2$, and $j$. Note, though,
that $\tilde\alpha = \alpha$, and $\tilde\beta = \beta$. Note also,
that we have inserted a  factor of $(1+\bxi_c
\xi_c)$ in the integral to compensate for the extra factor of
$(1+\bxi \xi)$ that was taken from the measure into the exponential 
to complete $\tilde S_{\rm tot}$. Thus {\it part\/} of both  the
measure and the prefactor are varied  in determining the stationary
phase, and get integrated over,  while part is regarded as a
constant.

The integration in Eq.~(\ref{EQ:Kcomb_j_tilde}) can be done at once
by noting that all equations in Appendix A are unchanged if we put
tildes on the actions, $j$, $\alpha$, and $\beta$ everywhere.
In particular, the identity (\ref{EQ:key_eqn2})
holds with tildes. We thus obtain
\be
K_{\rm comb}= \frac 1{\sqrt{2j+1}}
\left(\frac{\partial^2 \tilde S_{\rm tot}}{\partial \bzeta_f\partial\zeta_i}\right)^{\frac 12}
\exp\left\{\tilde S_{\rm tot}(\bzeta_f,\zeta_i,T)+\frac i2 \int_0^T Qdt \right\},
\ee
all unwanted factors of $2j+1$ and $(1+\bxi_c \xi_c)$, having
cancelled. Thus, with this form of stationary-phase 
integration, the propagator  reproduces itself exactly.

What this means is that the semiclassical approximation
must be  tacitly  using (\ref{EQ:2j+1decomposition}) in
making each of the many integrations that go into the
gaussian  approximation to the path integral. Once we
realize this, we see that there is no need for the 
mysterious divergent normalization factor, ${\cal N}=
\lim_{N\to\infty}(1+1/2j)^N$, that  plagues most
treatments of the semiclassical spin propagator.  

The appearance of $j+1/2$  as the large parameter in
the fluctuation integral has been remarked  on before   by
Ercolessi {\it et al.\/} \cite{ercolessi96} and by
Funahashi {\it et al.\/}\cite{funahashi95}. The former
worry that it is inconsistent to include fluctuations of 
the measure in the gaussian integral without also
considering their effect in the  saddle point equations. In
our case all terms that are being integrated over do appear
also in the equations determining the saddle point.

Note that the   correction $Q$ vanishes for 
Larmor
precession where $\hat H= \omega \hat J_3$.
In this case, as we have seen earlier, 
\be
S_{\rm cl}=2j\ln(1+ \bzeta_f \zeta_i e^{-i\omega T}) + i j\omega T.
\ee
 $\tilde S$ is obtained from this by the substitution $j\to j+\frac 12$, so
\be
\frac{\partial^2 \tilde S}{\partial \bzeta_f\partial\zeta_i}=
e^{-i\omega T} \frac {2j+1}{(1+\bzeta_f\zeta_ie^{-i\omega T})^2}.
\ee               
Thus
\bea
\frac 1{\sqrt{2j+1}}
\left(\frac{\partial^2 \tilde S}{\partial \bzeta_f\partial\zeta_i}\right)^{\frac 12}
e^{\tilde S(\bzeta_f,\zeta_i,T)}
&=&
e^{-i\omega T/2}
(1+\bzeta_f\zeta_ie^{-i\omega T})^{-1}
(1+\bzeta_f\zeta_ie^{-i\omega T})^{2j+1} e^{i\omega(j+\frac 12)T}\nonumber\\
&=&e^{ i\omega T}(1+\bzeta_f\zeta_ie^{-i\omega T})^{2j},
\eea   
which is the exact answer.

\section{An Example: \hbox{$\hH= \nu \hat J^2_3$}}

As an application of the semiclassical formalism 
consider    $\hH= \nu \hat J^2_3$.  This  hamiltonian is
time reversal invariant, and we might worry that a hidden
shift $j\to j+1/2$ would  compromise the Kramers degeneracy
expected when $j$ is half integral.

The  classical hamiltonian corresponding to $\hH= \nu \hat
J^2_3$  is
\be
H(\z,z)= {\eval{z}{\nu \hat J^2_3}{z} \over \brak{z}{z}}
       = \nu\left(j(j-\frac 12) 
            \left(\frac {\z z-1}{\z z+1}\right)^2
                +\frac 12 j\right).
\label{EQ:HJsq}
\ee
This should be compared with the ``na{\"\i}ve'' classical
hamiltonian
\be 
H_{\rm naive}=\nu j^2\left(\frac {\z z-1}{\z z+1}\right)^2,
\ee
which is what we would get if we simply expressed the
classical direction-dependent energy $\nu j^2
\cos^2 \theta$ in terms of the stereographic coordinates on $S^2$.

The hamiltonian (\ref{EQ:HJsq}) leads to the classical equations of motion
\be
\dot\z =i\omega(\z,z) \z, \quad \dot z = -i\omega(\z,z) z,
  \label{EQ:eom}
\ee
where, with $\mu = \nu j(j - 1/2)$, 
\be
\omega(\z,z)= \left(\frac {2\mu}{j}\right)\left(\frac {\z z-1}{\z z+1}\right).
\ee
Since these equations imply the time independence of the product $\z
z$, $\omega$ is itself time independent and  the solutions may be
written down directly as 
\be
z(t) = e^{-i\omega t}\zeta_i, \qquad \z(t) =e^{i\omega(t-T)}\bzeta_f.
\label{EQ:J2solutions}
\ee
Here $\omega$ is to be determined by the self-consistency condition
\be
\omega = \left(\frac {2\mu}{j}\right)
\left(\frac {e^{-i\omega T}\bzeta_f\zeta_i -1}{e^{-i\omega
T}\bzeta_f\zeta_i+1}\right).
  \label{EQ:omega_eqn}
\ee
As we will see below,  this equation has an infinite family of
solutions. Here, we wish to consider how various quantities scale
with $j$. By demanding that Eqs.~(\ref{EQ:eom}) continue to be
meaningful as $j \to \infty$, we see that we must have
$\mu = O(j)$, $\om = O(1)$, and $\nu = O(1/j)$.

The classical action for the solution (\ref{EQ:J2solutions}) is
\bea
S_{\rm cl}(\bzeta_f,\zeta_i,T)
    &=& 2j \ln(1+ e^{-i\omega T}\bzeta_f\zeta_i) \nonumber\\
    &&\quad +\int_0^T \left\{ j\left(\frac{ 2i\omega e^{-i\omega T}\bzeta_f\zeta_i}
       {1+e^{-i\omega T}\bzeta_f\zeta_i}\right) -i \mu \omega^2 
       \left(\frac {j}{2\mu}\right)^2 - {i\over 2}j \nu \right\}dt\nonumber\\
    &=& 2j \ln(1+ e^{-i\omega T}\bzeta_f\zeta_i) + 
       iT\{j\omega + \frac {j^2}{4\mu}\omega^2 -{1\over2}j\nu \}.
\label{EQ:J2action}
\eea
The apparently cosmetic rewrite in the last line leads to a useful way 
of looking at the problem. 
Define 
\be 
S_\omega(\bzeta_f,\zeta_i, T)=  2j \ln(1+ e^{-i\omega T}\bzeta_f\zeta_i) + 
iT\{j\omega + \frac {j^2}{4\mu}\omega^2 \},
\ee
where we regard $\omega$ as an independent variable. 
The equation 
\be
\frac {\partial S_\omega(\bzeta_f,\zeta_i,T)}{\partial \omega}=
iTj\left\{-\frac {e^{-i\omega T}\bzeta_f\zeta_i -1}{e^{-i\omega
T}\bzeta_f\zeta_i+1}+ \left(\frac {j}{2\mu}\right)\omega\right\}  
  \label{EQ:ptl_S_omega}
\ee
then shows that the consistency condition on $\omega$ is equivalent to 
${\partial S_\omega}/{\partial \omega}=0$. We can also use $S_\omega(\bzeta_f,\zeta_i,
T)$
to express the second variation of $S_{\rm cl}$ required for  the prefactor $A$.
By differentiating the  Jacobi equation
(\ref{EQ:jacobi_equations})  
we have 
\be
\frac{\partial^2 S_{\rm cl}(\bzeta_f,\zeta_i,T)}{\partial \bzeta_f\partial\zeta_i}
= \frac {2j}{(1+\bzeta_f z(T))^2}\frac{\partial z(T)}{\partial
\zeta_i},
\ee
and from this we find, with Eq.~(\ref{EQ:J2solutions}), that 
\be 
\frac{\partial^2 S_{\rm cl}(\bzeta_f,\zeta_i,T)}{\partial \bzeta_f\partial\zeta_i}
= 
  \frac {2j}{(1+\bzeta_f z(T))^2}\left\{ e^{-i\omega T} +  e^{-i\omega
T}\zeta_i \left( -iT \frac {\partial\omega}{\partial
\zeta_i}\right)\right\}.
 \label{EQ:action_ptls}
\ee
We now differentiate the condition $\ptl S_{\om}/\ptl \om = 0$
with respect to $\zeta_i$. This yields
\be
{\ptl^2 S_\om \over \ptl\zeta_i \ptl\om}
  + {\ptl^2 S_\om \over \ptl\om^2} {\ptl\om \over \ptl\zeta_i} = 0.
\ee
Using this result to eliminate $(\ptl\om / \ptl\zeta_i)$ in
Eq.~(\ref{EQ:action_ptls}), we find, after a little algebra, that
\be 
\frac{\partial^2 S_{\rm cl}(\bzeta_f,\zeta_i,T)}{\partial \bzeta_f\partial\zeta_i}
 = \frac {2je^{-i\omega T}}{(1+\bzeta_f z(T))^2}\cdot \frac
{iTj^2}{2\mu}\cdot \left(\frac{\partial^2 S_\omega}{\partial
\omega^2}\right)^{-1}.
   \label{EQ:action_ptls2}
\ee

Substituting Eqs.~(\ref{EQ:J2solutions}), (\ref{EQ:J2action}), and
(\ref{EQ:action_ptls2}) into the basic semiclassical form
(\ref{EQ:kochetov_propagator}) for the propagator, we obtain
\be
K_{\rm scl} = \sum_{\om} \left({iTj^2 \over 2\mu}\right)^{1/2}
   \left({\ptl^2 S_\om \over \ptl\om^2}\right)^{-1/2}
   \exp\left\{ S_\om - {iT\over 2}(\om + j\nu)
         + {i\over 2}\int_0^T A\,dt \right\}.
  \label{EQ:J2prop1}
\ee
The sum over $\om$ is to be performed over all solutions to
Eq.~(\ref{EQ:omega_eqn}).

The utility of $S_\omega(\bzeta_f,\zeta_i,T) $ is not hard to understand. 
We are trying to evaluate
\be
\eval{\zeta_f}{e^{-i\nu \hat J_3^2 T}}{\zeta_i}=
\sum_{m=-j}^{m=j} (\bzeta_f\zeta_i)^{j+m}
 \frac {2j!}{(j+m)!(j-m)!} e^{-i\nu m^2 T},  
\ee
while we already know that
\bea
\eval{\zeta_f}{e^{-i\omega \hat J_3T}}{\zeta_i}&=&
\sum_{m=-j}^{m=j} (\bzeta_f\zeta_i)^{j+m} \frac {2j!}{(j+m)!(j-m)!} e^{-i\omega m
T}\nonumber\\
&=& (1+ e^{-i\omega T}\bzeta_f\zeta_i)^{2j} e^{i\omega jT}\nonumber\\
&=& \exp S_{\omega 0}(\bzeta_f,\zeta_i,T),
\eea
where
\be
S_{\omega 0}(\bzeta_f,\zeta_i,\omega) =  2j \ln(1+ e^{-i\omega T}\bzeta_f\zeta_i) + 
iTj\omega.
\ee 

From the identity 
\be
e^{-i\nu m^2 T}= e^{-i\frac \pi 4}\sqrt{\frac{T}{4\pi\nu}} 
\int_{-\infty}^{\infty} d\omega  e^{-i\omega m T} e^{i\omega^2 T/4\nu}
\ee
we have the exact relation
\bea
\eval{\zeta_f}{e^{-i\nu \hat J_3^2T}}{\zeta_i}
&=&e^{-i\frac \pi 4}\sqrt{\frac{T}{4\pi\nu}} 
\int d\omega  \eval{\zeta_f}{e^{-i\omega \hat J_3}}{\zeta_i}
e^{i\frac{\omega^2 T}{4\nu}}
\nonumber\\
&=& e^{-i\frac \pi 4}\sqrt{\frac{T}{4\pi\nu}} 
\int d\omega  \exp\left\{S_{\omega 0}(\bzeta_f,\zeta_i,\omega) 
+i \frac {\omega^2 T}{4\nu}\right\} \nonumber\\
&=&e^{-i\frac \pi 4}\sqrt{\frac{T}{4\pi\nu}} 
\int d\omega\exp\left\{2j \ln(1+ e^{-i\omega T}\bzeta_f\zeta_i) + 
iT\{j\omega + \frac {\omega^2 }{4\nu}\}\right\}. 
\label{EQ:J2exactint}
\eea
Given the form of the classical action (\ref{EQ:J2action}), that
$\mu \approx j^2 \nu$,  and the occurrence of
$(\partial^2 S_\omega/\partial \omega^2)^{-1/2}$ in the prefactor,
it is clear that the semiclassical approximation  is attempting   
a stationary  phase
approximation to this integral over $\omega$.
That this approximation is indeed indicated can be seen by
evaluating $(\ptl^2 S_\om/\ptl \om^2)$. From
Eqs.~(\ref{EQ:ptl_S_omega}) and (\ref{EQ:omega_eqn}), we find
\be
{\ptl^2 S_\om \over \ptl\om^2}
  = {iTj^2 \over 2\mu}
      - {1\over 2}j T^2 \left( 1 - {j^2\om^2 \over 4\mu^2} \right),
\ee
which scales as $j$ as $j \to \infty$.

We now write the exponent in Eq.~(\ref{EQ:J2exactint}) as
$S_\om - iTj\om^2/8\mu$. Since the second term is $O(j^0)$
as $j \to \infty$, we may regard it as part of the pre-exponential
factor in carrying out the stationary phase integral. In this way,
we obtain
\be
K_{\rm exact} \approx
  \sum_\om \left( {iT\over 2\nu} \right)^{1/2}
    \left( {\ptl^2 S_\om \over \ptl\om^2} \right)^{-1/2}
    \exp\left\{ S_\om - iT {j\om^2 \over 8\mu} \right\}.
  \label{EQ:J2prop2}
\ee

The pre-exponential factors in the preceding equation agree with those
in Eq.~(\ref{EQ:J2prop1}) to terms of order unity. To see whether the
exponents agree, we must discuss
the effect of the Solari-Kochetov phase.
%At this point it is worth discussing the effect of the Solari-Kochetov phase.
We find that
\bea
A&=& \frac 12 \left(\frac{\partial}{\partial \z} \frac{(1+ \z z)^2}{2j}
\frac{\partial H}{\partial z} + \frac{\partial}{\partial z} \frac{(1+ \z z)^2}{2j}
\frac{\partial H}{\partial \z}\right)\nonumber\\
&=& \omega + \frac{4\mu}{j} \frac{\z z}{(1+\z z)^2} \nonumber\\
&=& \left( \om +{\mu\over j} \right) - {j\om^2 \over 4\mu}.
\eea
The term in parentheses serves to cancel [up to $O(1)$] the
second term in the exponent in Eq.~(\ref{EQ:J2prop1}), and the
$j\om^2/4\mu$ term serves to correct $S_\om$ as needed in
Eq.~(\ref{EQ:J2prop2}). Thus our semiclassical formula is indeed
accurate up to $O(1)$ as $j \to \infty$, and we may be confident
that spectral properties (Kramers degeneracy in particular) derived
from it by constructing, say, the Green's function or density of
states, will be faithfully given.

Having demonstrated the formal equivalence of $K_{\rm scl}$ and
$K_{\rm exact}$, we turn to the actual nature of the solution.
Let us first rewrite the self-consistency condition
(\ref{EQ:omega_eqn}) as
\be
2i \tom\tau + \ln \left( {1 + \tom \over 1 - \tom} \right)
  = \ln\alpha,
   \label{EQ:om_eqn2}
\ee
where $\tom = j\om/2\mu$, $\tau = \mu T/j$, and
$\alpha = {\bar\zeta}_f \zeta_i$. In the limit $\tau \to \infty$,
the left hand side of (\ref{EQ:om_eqn2}) must remain finite,
suggesting that $\tom \sim 1/\tau$. A development in powers
of $1/\tau$ shows that we may write
\be
\tom \approx -{i\over 2} {\ln\alpha \over \tau - i}
   - {1 \over 24}{ (\ln\alpha)^3 \over \tau^4}
   + O(\tau^{-5}).
  \label{EQ:om_tau_inf}
\ee
Since no restriction has been placed on which branch of $\ln\alpha$
is to be taken, this solution is infinitely multivalued, as asserted
above. To leading order in $1/\tau$, different solutions differ by additive
amounts $n\pi / \tau$, where $n$ is an integer.

On the other hand, at $\tau = 0$, Eq.~(\ref{EQ:om_eqn2}) has
a unique solution, $\tom = (\alpha - 1)/(\alpha + 1)$. The apparent
contradiction with the earlier argument for an infinite number of
solutions is resolved by noting that if, as $\tau \to 0$, we
allow $\tom$ to diverge as $1/\tau$, the left hand side of
(\ref{EQ:om_eqn2}) again remains finite. Another development in
powers of $\tau$ reveals that
\be
\tom \approx
    -{i\over 2} {\ln(-\alpha) \over \tau} - {2 \over \ln(-\alpha)}
     - {8i \over [\ln(-\alpha)]^3}\tau + \cdots,
  \label{EQ:om_tau_0}
\ee
which is also multivalued on account of the infinitely many branches of
$\ln(-\alpha)$. 

We can gain further insight into the nature of the propagator and the
values of $\omega$ at the relevant stationary-phase points by  
working with initial and final states on the equator of the sphere:
$\zeta_i=e^{i\phi_i}$, $\bzeta_f=e^{-i\phi_f}$. When $j$ is large,
the problem should be essentially equivalent to a massive particle
constrained to move on a ring of circumference $2\pi$. If we write
the hamiltonian for the latter as $L^2/2M$, where $L$ is the orbital
angular momentum, and $M$ the mass, we expect the results for the
two problems to be similar with $M = 2\nu$.

We start by considering the propagator for Larmor precession.
Employing  the leading large-$j$ estimate
\be
 \frac {2j!}{(j+m)!(j-m)!} \sim \frac{2^{2j}}{\sqrt{\pi j}}
e^{-m^2/j},
\ee
and 
using the shorthand 
$\Delta \phi = \phi_f-\phi_i$, 
we may write
\bea
\eval{\zeta_f}{e^{-i\omega \hat J_3}}{\zeta_i}
 &=& \sum_{m=-j}^{m=j} (\bzeta_f\zeta_i)^{j+m} \frac {2j!}{(j+m)!(j-m)!}
             e^{-i\omega m T} \nonumber\\  
&\sim& e^{-ij\Delta \phi}\frac{2^{2j}}{\sqrt{\pi j}} 
\sum_{m} e^{-im(\Delta \phi+\omega T)}e^{-m^2/j}.
\eea 
If $T \gg j^{-1/2}/\om$, the summand will have widely varying phases
over the range of $m$ values that contributes to the sum,
$|m| \sim \sqrt{j}$.
By extending the sum over $m$ to infinity and using the Poisson
summation formula (taking care that $m$ takes half-integer values when
$j$ is half integral), we find
\be
\eval{\zeta_f}{e^{-i\omega \hat J_3}}{\zeta_i}
\approx e^{-ij\Delta \phi} 2^{2j} \sum_n e^{-\frac {j}{4} (\Delta \phi
+\omega T -2\pi n)^2}\times (-1)^n,
\label{EQ:large_j_S_0}
\ee
where the $(-1)^n$ factor is present only when $j$ is half integral.
This form is better suited to studying the large $j$ limit (for fixed
$T$). In that case,
(\ref{EQ:large_j_S_0}), regarded as a function of
$\omega$,
is sharply peaked at $\omega=\bar \omega_n =(2\pi n -\Delta\phi)/T$. These 
are the angular frequencies that allow uniform precession
between $\phi_i$ and $\phi_f$ in time 
$T$. We now recall that Eq.~(\ref{EQ:large_j_S_0}) is nothing but
$\exp(S_{\om0})$. If we substitute this form into Eq.~(\ref{EQ:J2exactint}),
and take into account the factor $\exp\{i\omega^2 T/4\nu\}$ in determing the
saddle-point frequencies, we find that they become complex
\bea
\omega_n &=& \bar\omega_n\left(1 -\frac {i}{\nu T j}\right)^{-1}\nonumber\\
&\approx &  \bar\omega_n  + \frac {i\bar\omega_n}{\nu T j}.
\eea
Not surprisingly, this is just what we found in Eq.~(\ref{EQ:om_tau_inf}).
The result reflects the fact that, to move at the required speed, the 
hamiltonian trajectories must move off the  equator. There  is then
no real trajectory between the classical endpoints, and we must
exploit the freedom to have trajectories where $\bzeta_f \ne z(T)^*$.
When $j$ is large, however,  Hamilton's  equations provide large
%When $\nu$ is large, however,  Hamilton's  equations provide large
velocities  {\it close\/} to the equator, and the imaginary parts of
$\omega$ are correspondingly small. 
Performing the integration in Eq.~(\ref{EQ:J2exactint}), we find
\be
\eval{e^{i\phi_f}}{e^{-i\nu TJ^2_z}}{e^{i\phi_i}}\approx
2^{2j} e^{-ij\Delta\phi}\frac 1{(1+ij\nu T)^{\frac 12}}
\sum_n e^{-\frac j4  (\Delta \phi-2\pi n)^2/(1+ij\nu T) } 
\times (-1)^n,
  \label{EQ:sphereprop}
\ee
where, again, the last factor is only present when $j$ is
half-integral. This form should be compared with that for the
massive particle\cite{schulman81} 
\be
\eval{\phi_f}{e^{-iL^2 T/2M}}{\phi_i} =
{1 \over (2\pi i M T)^{1/2}}
  \sum_n \exp \left( in \Phi
              + i {M (\Delta\phi - 2 n\pi)^2 \over 2T} \right).
 \label{EQ:ringprop}
\ee
We have incorporated an Aharonov-Bohm phase $\Phi$ into the
result. This phase should be  $\pi$ when we compare with 
half-integer spins,
and the resulting pairwise degeneracy of the energy levels
is the particle-on-a-ring analogue of Kramers degeneracy.    

The similarity between Eqs.~(\ref{EQ:sphereprop}) and
(\ref{EQ:ringprop}) is evident.
Notice how $j$ sets the time scale for the crossover
between the large-$T$ regime, where the spin behaves essentially
as a particle of mass $2\nu$ on the ring,
and the short-time regime where the finite range of
the coherent-state
wavefunctions cuts off the $1/\sqrt T$ divergence.

Note that we have ignored the difference between $\mu/j^2$
and $\nu$ in the above comparison, since as discussed while
showing the equivalence of $K_{\rm scl}$ and $K_{\rm exact}$,
the error incurred is of order $1/j^2$ relative to the leading
term in the action. The semiclassical approximation therefore
correctly obtains the first two terms in the large-$j$
expansion.

\section{Discussion}

In the previous sections we have used the  continuous-time
path integral to  motivate  a   semiclassical approximation
to the coherent-state propagator for spin $j$.  Although
our derivation of the semiclassical propagator is purely
formal, and the resulting expression must initially have 
only the status of a conjecture, we   have demonstrated its
correctness by   verifying its short-time accuracy to
$O(T^2)$, and checking  its consistency under dissection of
the path. From these two properties we may conclude that
our expression is accurate to $O(j^0)$ uniformly in time.

In our derivation  it was necessary to take into account 
an ``anomaly'' in the evaluation of the functional
determinant of the Jacobi operator. This is the  only place
where we had to appeal to  details of the discrete version
of  the path integral. Regulating  the determinant  in a
manner consistent with the discrete  path integral  
results in a correction to the na{\"\i}ve expression for
the prefactor. This correction had been noted before, by
Solari\cite{solari87} and by Kochetov\cite{kochetov95}, but
its importance  does  not seem to have been widely
appreciated.

We have also discussed an example where an infinite number
of classical trajectories contribute to the propagator. Here
we again saw how the Solari-Kochetov factor is essential in 
obtaining the correct result. 

A calculation of the   Solari-Kochetov correction to   the
tunnel splitting between classically degenerate spin states
will be reported in a separate publication.

%\section{Acknowledgements}
\acknowledgements

Work at Urbana and Evanston was supported by the National Science
Foundation under grants DMR-98-17941 and DMR-9616749, respectively.
MS would also like to thank  the TCM
group at the Cavendish Laboratory, Cambridge, England, for
hospitality, and the  EPSRC for financial support under
grant number GR/N00364.

\appendix
\section{Composition of path-density factors} 
%\section{Derivation of (\ref{EQ:KEY_EQ})}

In the this appendix we derive (\ref{EQ:KEY_EQ}).
We begin by restating the stationary phase conditions
(\ref{EQ:stat_phase}):
\bea
0&=&\frac{\partial S_2(\bzeta_f,\xi)}{\partial \xi} - \frac{2j\bxi}{1+\bxi \xi},
\nonumber\\
0&=&\frac{\partial S_1(\bxi,\zeta_i)}{\partial \bxi} - \frac{2j\xi}{1+\bxi \xi}.    
\eea
Consider how the first of these evolves as we vary $\bzeta_f$. We find that 
\bea
0 &=&\frac{\partial}{\partial \bzeta_f}\left(\frac{\partial S_2}{\partial
\xi_c}- \frac{2j\bxi_c}{1+\bxi_c\xi_c}\right)\nonumber\\
&=&\frac{\partial^2 S_2}{\partial \bzeta_f\partial \xi_c}
+\frac{\partial^2 S_2}{\partial \xi_c^2}\frac{\partial\xi_c}{\partial \bzeta_f}
+ \frac {2j\bxi_c^2}{(1+\bxi_c\xi_c)^2}\frac{\partial\xi_c}{\partial \bzeta_f}
- \frac {2j}{(1+\bxi_c\xi_c)^2}\frac{\partial\bxi_c}{\partial \bzeta_f}\nonumber\\
&=& \frac{\partial^2 S_2}{\partial \bzeta_f\partial \xi_c}
+ \frac{\partial\xi_c}{\partial \bzeta_f}
\left(\frac{\partial^2 S_2}{\partial \xi_c^2}+  \frac
{2j\bxi_c^2}{(1+\bxi_c\xi_c)^2}\right) - 
\frac {2j}{(1+\bxi_c\xi_c)^2}\frac{\partial\bxi_c}{\partial \bzeta_f}.
\eea
In the last line, we recognize  the  expression in
parentheses to be
$2j\beta/((1+\bxi_c\xi_c)^2$, where $\beta$ is the coefficient appearing in 
(\ref{EQ:alpha_beta}). 
By differentiating each of the two stationary 
phase conditions with respect to $\bzeta_f$ and $\zeta_i$,
we get a total of four such equations. 
These  may be summarized as
\be
\left(\matrix{    1   &-\alpha\cr
               -\beta &    1  \cr}\right)
\left(\matrix{\frac{\partial \xi_c}{\partial \zeta_i}&\frac{\partial \xi_c}{\partial\bzeta_f}  \cr
              \frac{\partial \bxi_c}{\partial\zeta_i}&\frac{\partial \bxi_c}{\partial\bzeta_f}  \cr}
\right)  
=
\frac{(1+\bxi_c\xi_c)^2}{2j}
\left(\matrix{\frac{\partial^2 S_1}{\partial \bxi_c\partial\zeta_i}       & 0\cr
               0  & \frac{\partial^2 S_2}{\partial \bzeta_f\partial \xi_c}    
\cr}\right).
\label{EQ:matrix_relation}
\ee
Taking determinants, we obtain
\be
\left\vert\matrix{    1   &-\alpha\cr
               -\beta &    1  \cr}\right\vert
\left\vert\matrix{\frac{\partial \xi_c}{\partial \zeta_i}&\frac{\partial \xi_c}{\partial\bzeta_f}  \cr
              \frac{\partial \bxi_c}{\partial\zeta_i}&\frac{\partial \bxi_c}{\partial\bzeta_f}  \cr}
\right\vert  
=
\frac{(1+\bxi_c\xi_c)^4}{(2j)^2} \frac{\partial^2 S_1}{\partial \bxi_c\partial\zeta_i}     
                \frac{\partial^2 S_2}{\partial \bzeta_f\partial \xi_c}, 
   \label{EQ:det-product}
\ee

We now recall that the gaussian integration in Eq.~(\ref{EQ:redivide-K})
leads to the inverse-square root of the precisely the first determinant
in Eq.~(\ref{EQ:det-product}). This equation expresses this determinant in terms of
the second derivatives of $S_1$ and $S_2$, and the jacobian
$\ptl (\xi_c, {\bar\xi}_c)/\partial (\zeta_i, \bzeta_f)$.
The derivatives of $S_1$ and $S_2$
will cancel with the prefactors in Eq.~(\ref{EQ:redivide-K}), leaving only the
jacobian. We therefore turn to its evaluation, and show that it
can be written in terms of the second derivatives of $S_{\rm tot}$
with respect to $\bzeta_f$ and $\zeta_i$. We express $S_{\rm tot}$ as
\be
S_{\rm tot}=S_2+S_1-2j\ln(1+\bxi \xi),
\ee 
and take note of the fact that both $\xi_c$ and $\bxi_c$ vary as we vary
$\bzeta_f$ and $\zeta_i$.
We have
\bea
\frac{\partial^2 S_{\rm tot}}{\partial \bzeta_f\partial\zeta_i}&=&
\frac{\partial}{\partial\bzeta_f} \left(
\frac{\partial S_2}{\partial \xi_c}\frac{\partial \xi_c}{\partial \zeta_i}
-\frac {2j \bxi_c}{1+\bxi\xi}\frac{\partial \xi_c}{\partial \zeta_i}
+\frac{\partial S_1}{\partial \bxi_c}\frac{\partial \bxi_c}{\partial \zeta_i}
+ \frac{\partial S_1}{\partial \zeta_i}
-\frac {2j \xi_c}{1+\bxi\xi}\frac{\partial \bxi_c}{\partial
\zeta_i}\right)\nonumber\\
&=&\frac{\partial}{\partial\bzeta_f} \left(
\frac{\partial \xi_c}{\partial \zeta_i}\left\{\frac{\partial S_2}{\partial
\xi_c}-\frac {2j \bxi_c}{1+\bxi\xi}\right\}
+ \left(\frac{\partial S_1}{\partial \zeta_i}\right)_{\bxi_c}
+\left\{\frac{\partial S_1}{\partial 
\bxi_c}-\frac {2j \xi_c}{1+\bxi\xi}\right\}\frac{\partial \bxi_c}{\partial \zeta_i}
\right).
\eea 
The expressions in braces in the last line are the stationary phase conditions, so they are  zero, 
as are their derivatives. 
Thus:
\be
\frac{\partial^2 S_{\rm tot}}{\partial \bzeta_f\partial\zeta_i}=
\frac{\partial}{\partial\bzeta_f} 
\left(\frac{\partial S_1}{\partial \zeta_i}\right)_{\bxi_c}
=\frac{\partial^2 S_1}{\partial \bxi_c\partial\zeta_i}\frac{\partial \bxi_c}{\partial
\bzeta_f}.
\ee
Taking note of the fact that the derivative of $S_1$ with respect to $\zeta_i$ is at fixed $\bxi_c$, 
while  
we have useful expressions for the derivative including the variation of $\bxi_c$, we 
interchange the order of differentiation, and write 
\bea
\frac{\partial^2 S_{\rm tot}}{\partial \bzeta_f\partial\zeta_i}&=&
\frac{\partial \bxi_c}{\partial \bzeta_f}
\left(\frac{\partial}{\partial \zeta_i}\left(\frac{\partial
S_1}{\partial\bxi_c}\right)- 
\frac{\partial^2 S_1}{\partial \bxi_c^2} \frac{\partial \bxi_c}{\partial \zeta_i}\right)\nonumber\\
&=&
\frac{\partial \bxi_c}{\partial \bzeta_f}
\frac{\partial}{\partial \zeta_i}\left( \frac{2j\xi_c}{1+\bxi_c\xi_c}\right) -
\frac{\partial \bxi_c}{\partial \bzeta_f}\frac{\partial^2 S_1}{\partial \bxi_c^2} 
\frac{\partial \bxi_c}{\partial \zeta_i}\nonumber\\
&=&\frac{\partial \bxi_c}{\partial \bzeta_f}
\frac{\partial}{\partial \zeta_i}\left( \frac{2j\xi_c}{1+\bxi_c\xi_c}\right) -
\frac{\partial \bxi_c}{\partial \zeta_i}
\frac{\partial}{\partial \bzeta_f}\left(
\frac{2j\xi_c}{1+\bxi_c\xi_c}\right)\nonumber\\
&=&
\frac{2j}{(1+\bxi_c\xi_c)^2}\left(
\frac{\partial \bxi_c}{\partial \bzeta_f}\frac{\partial \xi_c}{\partial \zeta_i} 
-
\frac{\partial \bxi_c}{\partial \zeta_i}\frac{\partial \xi_c}{\partial
\bzeta_f}\right).
\eea
In going from the second line to the third, we used one of the
equations from (\ref{EQ:matrix_relation}).

Putting this together with (\ref{EQ:det-product}) yields 
\be 
\frac{\partial^2 S_{\rm tot}}{\partial \bzeta_f\partial\zeta_i}=
\frac{(1+\bxi_c\xi_c)^2}{2j}\frac{\partial^2 S_2}{\partial \bzeta_f\partial\xi_c}
 \frac{\partial^2 S_1}{\partial\bxi_c\partial\zeta_i}
\left\vert\matrix{ 1 & -\alpha\cr
               -\beta & 1 } \right\vert^{-1}
  \label{EQ:key_eqn2}
\ee
which is identical to Eq.~(\ref{EQ:KEY_EQ})

\eject

\begin{references}
%\begin{thebibliography}{99}

\bibitem{klauder79} J.~R.~Klauder, Phys. \ Rev.\ D {\bf 19} (1979) 2349.


\bibitem{kuratsuji80} H.~Kuratsuji, T.~Suzuki, J.\ Math.\  Phys.  {\bf 21}
(1980) 472.

\bibitem{jevicki79} A.~Jevicki, N.~Papanicolaou, Ann.\
Phys.\ (N.Y.) {\bf 120} (1979) 107.

\bibitem{rohrlich88} H.~B.~Nielsen, D.~R\"ohrlich, Nucl.\ Phys.\ B{\bf 299} 
(1988) 471.

\bibitem{klauder78} J.~R.~Klauder, in {\it Path Integrals\/} Proceedings of the NATO
Advanced Summer Institute, edited by G.~J.~Papadopoulos and
J.~T.~Devreese (Plenum, NY ,1978).

\bibitem{inomata92}  {\it Path integrals and coherent
states of $SU(2)$ and $SU(1,1)$\/}, H. Kuratsuji,  A. Inomata, C.~C.~Gerry,  eds.
(World Scientific, Singapore, 1992).

\bibitem{woodhouse} N.~M.~J.~Woodhouse, {\it Geometric Quantization\/}
(Oxford University Press 1992).

\bibitem{duistermat82} J.~H.~Duistermat, G.~J.~Heckman, Invt.\ Math.\ {\bf 69} (1982)
259; {\bf 72}(1983) 153.

\bibitem{stone89} M.~Stone, Nucl.\ Phys.\ B {\bf 314} (1989) 577.

\bibitem{semenoff91} E.~Keskivakkuri, A.~J.~Niemi,
G.~Semenoff, O.~Tirkkonen, Physical Review {\bf D 44}
(1991) 3899.

\bibitem{chudnovsky88} E.~M.~Chudnovsky, L.~Gunther, Phys.\  Rev.\  Lett.\ {\bf
60} (1988) 661.

\bibitem{loss92} D.~Loss, D.~P.~Vincenzo, G.~Grinstein,
Phys.\ Rev.\ Lett.\ {\bf 69} (1992) 3232.

\bibitem{henley92} J.~von Delft and C.~L. Henley, 
Phys.\ Rev.\ Lett.\ {\bf 69} (1992) 3236.

\bibitem{garg93} A.~Garg, Europhys.\ Lett.\ {\bf 22}, 205 (1993). 

\bibitem{wernsdorfer99} W.~Wernsdorfer and R.~Sessoli,
Science {\bf 284}, 133 (1999).

\bibitem{kuratsuji81} H.~Kuratsuji, Y.~ Mizobuchi, J.\ Math.\  Phys.  {\bf 22}
(1981) 757.    

\bibitem{garg_kim92} A.~Garg, G-H.~Kim, Phys.\ Rev.\ B {\bf 45} (1992)
921.

\bibitem{enz86} M.~Enz, R.~Schilling, J.\ Phys.\  C {\bf 19} (1986) L711;
1765.

\bibitem{belinicher97} V.~I.~Belinicher, C.~Providencia, J.~da~Providencia, 
J.\ Phys.\ A {\bf 30} (1997) 5633.

\bibitem{shibata99} 
J.~Shibata, S.~Takagi,  Int.\ Jour.\   Mod.\  Phys.  B {\bf
13} (1999) 107.

\bibitem{hemmen86} J. L Hemmen A.~S\"ut{\H o}, Europhys.\ Lett.\ {\bf 1} (1986) 481; 
Physica B \& C {\bf 141} (1986) 37.

\bibitem{garg99} A. Garg,
Phys.\ Rev.\ Lett.\ {\bf 83}, 4385 (1999); cond-mat/0003114;
cond-mat/0003156.

\bibitem{villain00} J.~Villain and A.~Fort, submitted to Europhys.\ B.

\bibitem{solari87} H.~G.~Solari,  J.\  Math.\ Phys.\ {\bf 27} (1987)
1097.

\bibitem{kochetov95} E.~A.~Kochetov, J.\  Math.\ Phys.\ {\bf 36} (1995)
4667.

\bibitem{perelomov86} A.~Perelomov, {\it Generalized Coherent
States and their Applications\/}, (Springer-Verlag, Berlin 1986).

\bibitem{borel}See for example: R.~Bott, Ann.\ Math.\ {\bf 66} (1957)
203.

\bibitem{kirillov} A.~A.~Kirillov, {\it Elements of the Theory of
Representations\/}, (Springer-Verlag, Berlin, New-York, 1976).

\bibitem{lieb73} E.~Lieb, Comm.\ Math.\ Phys.\ {\bf 31}
(1973) 327.

\bibitem{richtmeyer78}See for example: R.~D.~Richtmyer, 
{\it Principles of Advanced Mathematical Physics\/}, 
(Springer Verlag, Berlin, New-York,  1978), section 7.2.

\bibitem{balachandran83} A.~P.~Balachandran, G.~Marmo, B.~S.~Skagerstam, A.~Stern, {\it 
Gauge Symmetries and Fibre Bundles\/}, (Springer-Verlag, Berlin, 1983);
 A.~P.~Balachandran, G.~Marmo, B.~S.~Skagerstam, A.~Stern, {\it
Classical topology and Quantum States\/}, (Springer-Verlag, Berlin, 1991). 

\bibitem{stone-kos98} The ``shooting method'' we used is  described in: 
S.~Kos, M.~Stone, Phys.\ Rev.\ B {\bf
59} (1999) 9545. (cond-mat/9809182)

\bibitem{marinov80} M.~S.~ Marinov, Phys.\ Rep.\ {\bf 60}
(1980) 1.

\bibitem{ercolessi96}  E.~Ercolessi, G.~Morandi, F.~Napoli, R.~Pieri, J.\ Math.
Phys.\ {\bf 37} (1996) 535.

\bibitem{funahashi95} F.~Funahashi, T.~Kashiwa, S.~Sakoda, K. Fujii, J.\ Math.
Phys.\ {\bf 36} (1995) 3232.

\bibitem{schulman81} See, e.g., L.~S.~Schulman, {\it Techniques and
Applications of Path Integration\/},  (Wiley, New York, 1981),
Chapter. 23.










%\end{thebibliography} 
\end{references}
\end{document}